\documentclass[sigconf]{acmart}
\usepackage{cleveref}
\usepackage{bm}
\usepackage{bbm}
\usepackage{caption}
\usepackage{subcaption}
\usepackage{booktabs,amsfonts,dcolumn}
\usepackage{mathtools}
\usepackage{wasysym}
\usepackage{hhline}
\usepackage{float}
\usepackage{commath}

\DeclareMathOperator{\E}{\mathbb{E}}
 \usepackage{multirow}
 \usepackage[ruled,vlined,linesnumbered]{algorithm2e}
\usepackage{booktabs,siunitx}

\usepackage{stmaryrd}
\usepackage{trimclip}
\usepackage{pifont}
\usepackage{enumitem}
\usepackage{mleftright}
\usepackage{scalerel,stackengine}
\stackMath
\newcommand\reallywidehat[1]{%
\savestack{\tmpbox}{\stretchto{%
  \scaleto{%
    \scalerel*[\widthof{\ensuremath{#1}}]{\kern-.6pt\bigwedge\kern-.6pt}%
    {\rule[-\textheight/2]{1ex}{\textheight}}%WIDTH-LIMITED BIG WEDGE
  }{\textheight}% 
}{0.5ex}}%
\stackon[1pt]{#1}{\tmpbox}%
}

\usepackage{titlesec}
\AtBeginDocument{%
  \providecommand\BibTeX{{%
    \normalfont B\kern-0.5em{\scshape i\kern-0.25em b}\kern-0.8em\TeX}}}

\copyrightyear{2023}
\acmYear{2023}
\setcopyright{rightsretained}
\acmConference[WSDM '23]{Proceedings of the Sixteenth ACM
International Conference on Web Search and Data Mining}{February
27-March 3, 2023}{Singapore, Singapore}
\acmBooktitle{Proceedings of the Sixteenth ACM International Conference
on Web Search and Data Mining (WSDM '23), February 27-March 3, 2023,
Singapore, Singapore}
\acmDOI{10.1145/3539597.3570474}
\acmISBN{978-1-4503-9407-9/23/02}

\begin{document}

\title{Marginal-Certainty-aware Fair Ranking Algorithm}

\author{Tao Yang}

\affiliation{%
	\institution{University of Utah}
	\streetaddress{50 Central Campus Dr.}
	\city{Salt Lake City}
	\state{Utah}
	\country{USA}
	\postcode{84112}
}
\email{taoyang@cs.utah.edu}

\author{Zhichao Xu}
\affiliation{%
	\institution{University of Utah}
	\streetaddress{50 Central Campus Dr.}
	\city{Salt Lake City}
	\state{Utah}
	\country{USA}
	\postcode{84112}
}
\email{zhichao.xu@utah.edu}

\author{Zhenduo Wang}
\affiliation{%
	\institution{University of Utah}
	\streetaddress{50 Central Campus Dr.}
	\city{Salt Lake City}
	\state{Utah}
	\country{USA}
	\postcode{84112}
}
\email{zhenduow@cs.utah.edu}
\author{Anh Tran}
\affiliation{%
   \institution{University of Utah}
	\streetaddress{50 Central Campus Dr.}
	\city{Salt Lake City}
	\state{Utah}
	\country{USA}
	\postcode{84112}
}
\email{abtran@cs.utah.edu}

\author{Qingyao Ai}
\affiliation{%
   \institution{Tsinghua University}
     \streetaddress{50 Central Campus Dr.}
   \city{Beijing}
   % \state{China}
   \country{China}
   \postcode{100084}
}
\email{aiqy@tsinghua.edu.cn}

\begin{abstract}
% Ranking systems are the main interface where online platforms like recommender system and search engine provide a list of items to users. 
Ranking systems are ubiquitous in modern Internet services, including online marketplaces, social media, and search engines.
Traditionally, ranking systems only focus on how to get better relevance estimation. 
When relevance estimation is available, they usually adopt a user-centric optimization strategy where ranked lists are generated by sorting items according to their estimated relevance. 
However, such user-centric optimization ignores the fact that item providers also draw utility from ranking systems. 
It has been shown in existing research that such user-centric optimization will cause much unfairness to item providers, followed by unfair opportunities and unfair economic gains for item providers. 

% The unfairness will force providers to leave the system, and discourage new providers from coming in. Finally, fewer purchase options would be left for customers and ruin the ranking system in the long run.
To address ranking fairness, many fair ranking methods have been proposed. 
However, as we show in this paper, these methods could be suboptimal as they directly rely on the relevance estimation without being aware of the uncertainty (i.e., variance of the estimated relevance).
To address this uncertainty, we propose a novel \textbf{M}arginal-\textbf{C}ertainty-aware \textbf{Fair} algorithm named \textbf{MCFair}. 
MCFair jointly optimizes fairness and user utility, while relevance estimation is constantly updated in an online manner. 
% To address the uncertainty, we propose MCFair, a Marginal-Certainty-aware Fair algorithm which optimizes fairness and user utility while relevance estimation is constantly updated in an online manner. 
In MCFair, we first develop a ranking objective that includes uncertainty, fairness, and user utility. 
Then we directly use the gradient of the ranking objective as the ranking score. 
We theoretically prove that MCFair based on gradients is optimal for the aforementioned ranking objective. % Zhenduo
Empirically, we find that on semi-synthesized datasets, MCFair is effective and practical and can deliver superior performance compared to state-of-the-art fair ranking methods. To facilitate reproducibility, we release our code.\footnote{https://github.com/Taosheng-ty/WSDM23-MCFair}
\end{abstract}






\maketitle

\section{INTRODUCTION}
Advanced ranking techniques have led to improvements in AI-powered information services that significantly changed people’s lives. For example, search engines that rank documents according to their utilities to user’s queries have helped billions of people better finish their daily work; recommendation systems that rank products/movies/news according to the user's interests have completely changed the way people obtain information every day. Therefore, how to construct and optimize ranking systems is one of the crucial research problems in Information Retrieval (IR)~\cite{liu2009learning}.

When considering the quality of result rankings in a ranking system, there are two important criteria: \textit{Ranking Effectiveness} and \textit{Ranking Fairness} \cite{morik2020controlling,singh2018fairness,biega2018equity}.
Ranking Effectiveness refers to the ability of a ranking system to effectively put relevant results at the top ranks;
by maximizing ranking effectiveness, we can help save users' efforts as they only need to examine the top ranks to satisfy their information needs \cite{wang2018position,joachims2017accurately}.
However, myopically optimizing ranking effectiveness according to relevance can lead to unfair ranking results.
For example, in a hiring website, if a ranking system only considers ranking effectiveness and ranks candidates solely according to relevance, then a small number of top candidates will always be exposed to employers and dominate employers' attention as employers usually only examine the top ranks \cite{joachims2017accurately}. In this case, other candidates will be unfairly treated and rarely have the chance to be hired even when they are also highly qualified for the job. 
Therefore, it is critical to jointly consider ranking fairness and ranking effectiveness.
Formally, ranking fairness measures the ranking system's ability to present fairly \cite{singh2018fairness}.
In this work, we focus on the important problem of \textit{Exposure Fairness}, as exposure directly influences opinion (e.g., ideological orientation of presented news articles) or economic gain (e.g., revenue from product sales or streaming) for providers of items~\cite{morik2020controlling}. 
Relevance estimation serves as the foundation of the optimization of effectiveness and fairness.
Specifically, optimizing effectiveness means putting more relevant items on top ranks, while optimizing fairness means letting items of similar relevance receive similar exposure.
To jointly optimize ranking effectiveness and fairness, many fair ranking methods \cite{singh2018fairness,biega2018equity,yang2022effective} adopt a post-processing setting that assumes that relevance is well estimated prior to the effectiveness-fairness joint optimization. However, such a post-processing setting seldom exists.
% and there are at least two challenges that require careful design consideration in a real world scenario.
In a real-world scenario, relevance estimation and ranking optimization are usually dynamically entangled with each other in an online way. 
Relevance estimation influences how the ranked lists are optimized, and the ranked lists will be later presented to users to collect their feedback, which will, in return, influence relevance estimation. % Zhenduo
From a statistical point of view, relevance estimation usually comes with uncertainty, i.e., variance  and relevance estimations for different items are not equally trustworthy since their uncertainty is usually not the same. 
Optimizing ranking effectiveness and fairness without considering such differences in uncertainty will make existing post-processing fair methods suboptimal. 
It has come to our notice that although some methods~\cite{morik2020controlling,yang2021maximizing} have been proposed to balance ranking effectiveness and fairness in an online setting, these methods overlook the uncertainty difference in relevance estimation. % Zhenduo
In this paper, we will show these uncertainty-oblivious ranking methods are suboptimal in the online setting.

% Given the interplay between relevance estimation and ranking optimization, jointly optimizing effectiveness and fairness needs careful design consideration since
% From the point of view of statistics, relevance estimation comes with an uncertainty, i.e, variance.
% relevance estimations for different items are not equally trustworthy since their uncertainty are usually not the same. 
% Second, the partial-information nature of user feedback will introduce bias in relevance estimation. For example, items ranked highly are more likely to gain additional feedback, such as clicks, due to the position bias~\cite{joachims2017accurately}, but are not necessarily  more relevant. Items haven't been presented to users before may never collect clicks, but are not necessarily irrelevant , which is the so-called  the selection bias~\cite{oosterhuis2020policy}. Failing to estimate relevance well will result in a failure to ranking optimization. Although \citet{morik2020controlling} considers the position bias, we found, in this paper, that the existence of selection bias makes their method fail. 

In this paper, we propose a \textbf{M}arginal-\textbf{C}ertainty-aware \textbf{Fair} ranking algorithm, or \textbf{MCFair} to jointly optimize effectiveness and fairness in an online setting. 
This algorithm addresses the dynamic nature of online setting where rank optimization is carried out while the relevance is still being learnt from users' biased feedback. 
The core of our algorithm is to first formulate a ranking objective that includes effectiveness, fairness, and uncertainty, then take derivatives of the ranking objective with respect to exposure and directly use the gradients as ranking scores. % Zhenduo
The ranking scores from the gradients automatically include a marginal-certainty-aware exploration strategy to deal with the uncertainty in relevance estimation. 
We theoretically prove that the ranking scores, i.e., gradients, are optimal for the ranking objective.  
In addition to the theoretical justification of MCFair, we provide empirical results with two real-world datasets under both the post-processing setting and the online setting. We find that MCFair outperforms existing state-of-the-art methods significantly. Furthermore, MCFair is efficient, robust, and easy to implement. 

% \brutuscomment{TODO}

\section{RELATED WORK}

\textbf{\textit{Exposure Fairness in Ranking.}}
Fairness has been a heated research topic in the IR community, especially in ranking \cite{zehlike2017fa,zehlike2022fair,zehlike2020reducing,oosterhuis2020policy,morik2020controlling},
There exist various definitions and criteria of ranking fairness~\cite{mehrabi2021survey,singh2021fairness}. 
% such as \textit{individual fairness}, i.e. to treat similar individuals similarly \cite{dwork2012fairness}, and \textit{groupfairness}, i.e. to treat different groups of individuals defined by certain characteristics such as demographics should be treated in a similar manner \cite{mehrabi2021survey,speicher2018unified}. 
In this work, we specifically focus on \textit{Exposure Fairness}, which is crucial for ranking services. For exposure fairness, 
\citet{biega2018equity} and \citet{singh2018fairness} independently propose the well-known amortized fairness principle where items' exposure should be proportional to their relevance.
Some prior works \cite{yang2021maximizing,yang2022effective,biega2018equity,singh2018fairness,singh2019policy,morik2020controlling,wu2021tfrom,oosterhuis2021computationally} have been proposed to address the amortized fairness.
Among these works, \cite{yang2021maximizing,morik2020controlling} choose to first identify unfairly-treated items based on some predefined metrics, then boost these items' ranking scores to gain more exposure and mitigate existing unfairness.
However, these methods based on predefined metrics could be suboptimal because they don't directly optimize fairness (more discussion in \S\ref{sec:gradientBased}). 
Another line of works \cite{biega2018equity,singh2018fairness} proposes to use linear programming (LP) methods to directly optimize fairness.
However, the number of decision variables of these LP methods is $O(n^2)$, which quadratically increases when the number of candidate items \textit{n} increases. 
This $n$ can be a huge number in real-world ranking applications, and $O(n^2)$ decision variables become the bottleneck of the LP methods.

\textbf{\textit{Uncertainty in Ranking.}}
Model Uncertainty has been widely studied in ML and Statistics community \cite{draper1995assessment,gal2016dropout}.
In IR literature, Zhu et al. \cite{zhu2009risk} propose to improve retrieval performance by utilizing the variance of a probabilistic language model as the risk-based factor.
Some other works also make use of model uncertainty to improve query performance prediction \cite{roitman2017robust,schuth2015probabilistic}, query cutoff prediction \cite{culpepper2016dynamic,lien2019assumption} as well as neural IR \cite{penha2021calibration,cohen2021not}.
However, the exploration of uncertainty in learning to rank has become popular recently. For example,
\citet{singh2021fairness} proposed a new type of ranking fairness based on uncertainty.
A most recent work from \citet{yang2022CanClicks} proposes to use uncertainty to explore cold-start items in online ranking services.  In this paper, unlike previous works which use uncertainty directly, we propose to use marginal (un)certainty to guide exploration.
% \brutuscomment{how is this work related to uncertainty?}

\textbf{\textit{Unbiased and Online LTR.}}
Unbiased learning to rank focuses on training unbiased models with biased click signals \cite{ai2018unbiased,ai2021unbiased,joachims2017accurately}, which have been a heated topic in recent years~\cite{tranutirl,zhao2022overview}.
Offline LTR trains the unbiased model with offline click logs \cite{ai2018learning,joachims2017accurately,oosterhuis2021unifying,luo2022model,tran2021ultra,yang2020analysis} while online LTR focuses on actively removing biases from labels via online interpolations \cite{yang2022CanClicks,hofmann2013balancing}.
Within online LTR, some works propose to explore the relevance of result rankings with bandit learning \cite{schuth2016multileave,wang2019variance,yue2009interactively,xu2022reinforcement}; 
\citet{oosterhuis2020policy} propose to dynamically estimate items' relevance via stochastic ranking sampling. 
\citet{yang2022CanClicks} analyze the effect of using behavior features (click data) as both features and labels for online LTR. 
However, most of the existing unbiased and online LTR methods only focus on estimating relevance and maximizing effectiveness. In this paper, following~\cite{morik2020controlling}, we will explore an algorithm to jointly optimize ranking effectiveness (relevance) and fairness.

% \subsection{Challenges in optimization}
% \label{sec:challenges}
% In the optimization part of in Figure~\ref{fig:workflow}, there exist two challenges. One is the optimization itself and the other is brought by the relevance estimation based on users feedback. 
% In this section, we will introduce the two challenges. 

% \subsubsection{Challenge in Joint Optimization.} The first challenge is how to design algorithm to jointly optimize effectiveness and fairness.

% Some existing works ~\cite{morik2020controlling,yang2021maximizing} choose to first identify unfairly treated items based on some heuristic judgement and then boost the ranking scores of unfairly treated items to mitigate unfairness. Those heuristic methods work but mostly not optimal as they don't directly optimize the objectives, which we will show in our experimental results.
% Besides those heuristic methods, another group of methods use linear programming ~\cite{singh2018fairness,biega2018equity} to directly optimize fairness. However, the number of decision variables of those linear programming methods is $O(n^2)$ which qudratically increases as the number of candidate items, $n$, for a query increases. Since the number of decision variables is the bottleneck of linear programming, those methods are seldom used in real-world applications since  $n$ is mostly a large number. 
\section{BACKGROUND}
In this section, we give the background knowledge for this paper, which includes the workflow of ranking service (\S\ref{sec:workflow}), metrics for ranking optimization (\S\ref{sec:rankingutilitymeasurement}) and biased user feedback in ranking service (\S\ref{sec:biasedFeedback}). A summary of notations is shown in Table \ref{tab:notation}.

\begin{figure}[t]
    \centering
    \caption{Workflow of ranking services.}
    \includegraphics[width=1.0\columnwidth]{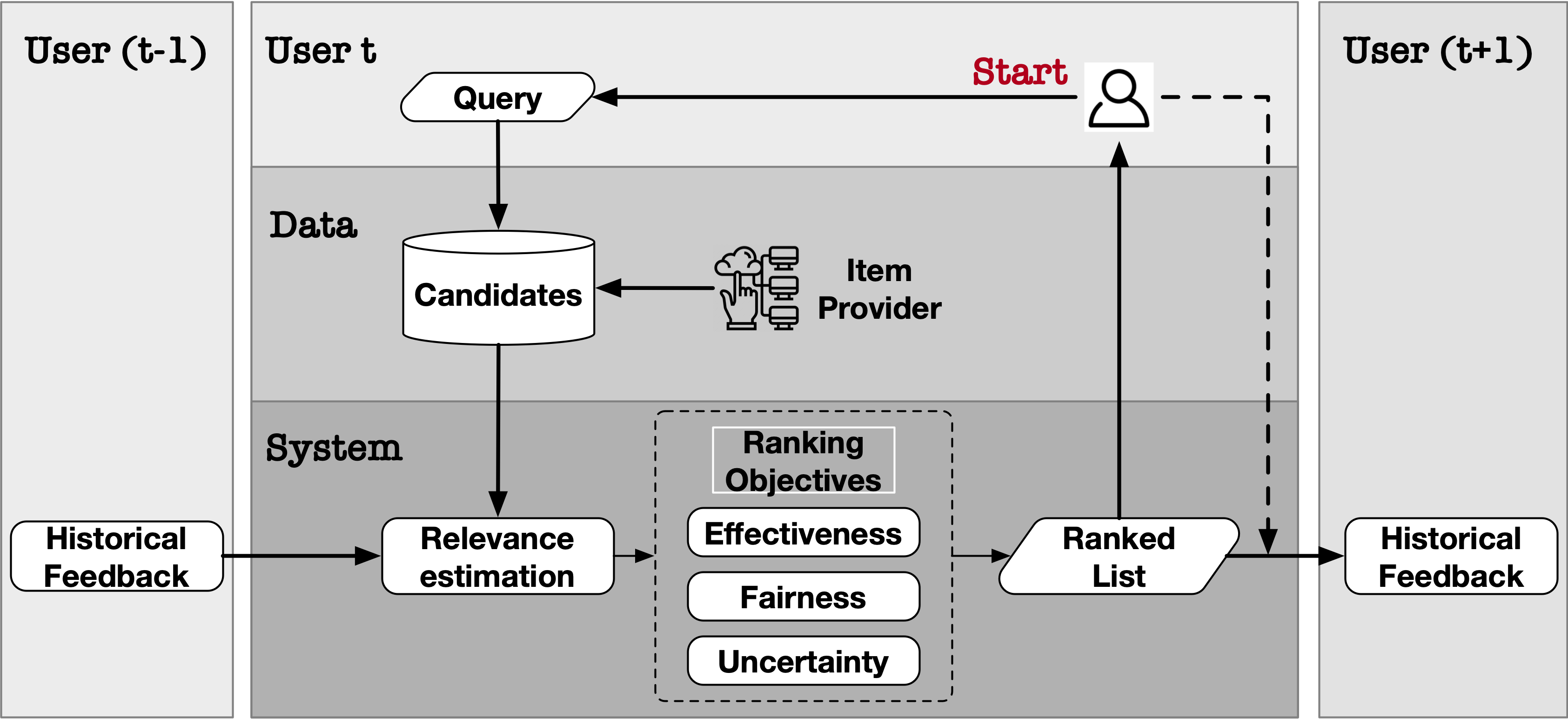}
    \label{fig:workflow}
\end{figure}

\subsection{The Workflow of Ranking Service}
\begin{table}[t]
\vspace{-15pt}
	\caption{A summary of notations.}
    \vspace{0pt}
	\small
	\def\arraystretch{1}%  1 is the default, change whatever you need
	\begin{tabular}
		{| p{0.07\textwidth} | p{0.36\textwidth}|} \hline
		$d,q,D(q)$ & For a query $q$, $D(q)$ is the set of candidates items. $d\in D(q)$ is an item. \\\hline
		$e,r,c$ & All are binary random variables indicating whether an item $d$ is examined ($e=1$), perceived as relevant ($r=1$) and clicked ($c=1$) by a user respectively. \\\hline
		$R,p_i,E,\pi$ & $R=p(r=1|d)$,  is the probability of an item $d$ perceived as relevant. $p_i=p(e=1|rnk(d)=i,\pi)$ is the examination probability of item $d$ when it is put in $i^{th}$ rank in a ranklist $\pi$. $E$ is item's accumulated examination probability (see Eq.\ref{eq:expo}).  \\\hline
		$k_s,k_c$ & Users will stop examining items lower than rank $k_s$ due to selection bias (see Eq.~\ref{eq:OverallBias}).  $k_c$ is the cutoff prefix to evaluate cNDCG and $k_c\le k_s$.\\\hline
	\end{tabular}\label{tab:notation}
\vspace{0pt}
\end{table}

\label{sec:workflow}

Here we introduce the workflow of ranking service in detail. 
Figure~\ref{fig:workflow} is the flowchart of a ranking task. 
At time step $t$, a user issues a query, and there are several candidate items corresponding to this query. 
Then the relevance estimator predicts the relevance of each candidate item and the ranking optimization methods will generate the ranked list by optimizing the ranking objective based on relevance estimation. 
% Different ranking objectives can be adopted here.  %Zhenduo
There are a few different ranking objectives can be adopted here. 
% For example, the ranking objective can be to maximize effectiveness and fairness while minimizing the uncertainty. 
For example, the ranking objective can be to jointly optimize effectiveness and fairness \cite{yang2022effective}.
%After examining the ranked list, the user will provide her feedback, such as clicks. With user's feedback, the relevance estimator will update relevance estimation for future ranking optimization.  %Zhenduo
After examining the ranked list, the user will provide their feedback, such as clicks. With user's feedback, the relevance estimator will update relevance estimation for future ranking optimization.  
The generated ranked list will be evaluated by the ranking utility measurement.
% Ranking utility measurement will evaluate the generated ranked list.

\subsection{Ranking Utility Measurement}
\label{sec:rankingutilitymeasurement}
Ranking is a two-sided market, from which users and item providers both draw utility.
Here we introduce the concepts of user-side utility and provider-side utility, which provide guidance on how we optimize and evaluate ranked lists.
\subsubsection{The User-side Utility (Effectiveness).} 
Before introducing the user-side utility, we define the relevance $R(d,q)$ in this paper
as the probability of an item $d$ to be relevant to query $q$:
\begin{equation}
\vspace{0pt}
    R(d,q)=P(r=1|d,q)
\vspace{0pt}
\end{equation}
where $r$ is a binary random variable indicating $d$ perceived by a user as relevant to query $q$ or not.

As users are the main clients of ranking systems, it is important to evaluate ranking performance from the user side. 
The user-side utility, also referred to as \textit{effectiveness}, is usually used to measure a ranking system's ability to put relevant items on top ranks. One widely-used user-side utility measurement is Discounted Cumulative Gain~\cite{jarvelin2002cumulated}, denoted as $\textit{DCG}$. For a ranked list $\pi$ corresponding to a query $q$, we define $\textit{DCG}@k_c$ as:
\begin{equation}
\vspace{0pt}
    \textit{DCG}@k_c(\pi)=\sum_{i=1}^{k_c}R(\pi[i],q)\lambda_i
    \label{eq:DCG}
\vspace{0pt}
\end{equation}
where $\pi[i]$ indicates the $i^{th}$ ranked item in the ranked list $\pi$; $R(\pi[i],q)$ indicates item $\pi[i]$'s relevance to query $q$; cutoff $k_c$ indicates the top ranks we evaluate; $\lambda_i$ indicates the weight we put on $i^{th}$ rank.  
$\lambda_i$ usually monotonically decreases as rank $i$ increases since top ranks are generally more important. 
For example, $\lambda_i$ is sometimes set to $\frac{1}{\log_2(i+1)}$. 
In this paper, we follow \cite{singh2018fairness} and set $\lambda$ as the examining probability $p$:
% \begin{equation}
% \begin{split}
%     \textit{eff.}(q,t)&=\sum_{i=1}^t \sum_{j=1}^k p_{j} R(\pi_i[j])
%     \end{split}
% \label{eq:cumDCG}
% \end{equation}
\begin{equation}
\vspace{0pt}
    \textit{DCG}@k_c(\pi)=\sum_{i=1}^{k_c}R(\pi[i],q) \cdot p_i
    \label{eq:DCG}
\vspace{0pt}
\end{equation}
where $p_{i}$ is users' examining probability of the $i^{th}$ item in ranked list $\pi$, and we have:
\begin{equation}
\vspace{0pt}
    p_{i}=P(e_{i}=1)
% \vspace{0pt}
\end{equation}
where $e_{i}$ is a binary variable indicating the $i^{th}$ item being examined  or not. 
% It is straightforward that we can get the ideal $DCG$ ($iDCG$), the maximum $DCG$, by ranking items according to relevance.
Then, we can define the normalized-$\textit{DCG}$ ($\textit{NDCG}$) by normalizing $\textit{DCG}@k_c(\pi)$ with  $\textit{DCG}@k_c(\pi^*)$:
\begin{equation}
\vspace{0pt}
    \textit{NDCG}@k_c(\pi)=\frac{\textit{DCG}@k_c(\pi)}{\textit{DCG}@k_c(\pi^*)}
    \label{eq:NDCG}
\vspace{0pt}
\end{equation}
where $\pi^*$ is the ideal ranked list constructed by arranging items by their true relevances and $\textit{NDCG}@k_c(\pi) \in [0,1]$.
Furthermore, we could define Cumulative NDCG (cNDCG) as:
\begin{equation}
\vspace{0pt}
\begin{split}
   \textit{eff.}= \textit{cNDCG}@k_c=\sum_{\tau=1}^t \gamma^{t-\tau} \textit{NDCG}@k_c(\pi_\tau)
    \end{split}
\label{eq:cumDCG}
\vspace{0pt}
\end{equation}
where $0\le\gamma\le 1$ is the discounted factor, $t$ is the current time step. Note that $\gamma$ is usually set as a constant for all time steps.
Compared to NDCG, cNDCG can better evaluate effectiveness for online ranking services ~\cite{schuth2013lerot}. 
Since we mainly consider online ranking (as shown in Figure~\ref{fig:workflow}), \textbf{we use cNDCG as the effectiveness objective and measurement in this work.}
% \brutuscomment{not sure what's the exact connection here}

% As $\lambda_i$ usually monotonically decreases as $i$ increases, It is straightforward that we can reach the maximum $\textit{NDCG}$ by ranking items according to relevance in descending order,
% \begin{equation}
% 	\pi_t=\arg_{topk}  \big(\hat{R}(d,q_t), \forall d\in D_{q_t}\big)
% 	\label{equ:score_rank}
% \end{equation}
% where $\hat{R}(d,q_t)$ is item $d$'s relevance score  predicted by ranking models for query $q_t$ at time step $t$, $k$ is the length of the returned list to users. We refer to this ranking strategy as TopK.

\subsubsection{The Provider-side Utility (Fairness).} 
Since items' rankings can determine their providers' utility, it is also important to evaluate ranking performance from the provider perspective. 
% Ranking fairness is often chosen as the provider side utility that evaluate ranking systems' ability to create a fair environment for items and their providers\footnote{In this paper, we use item fairness and provider fairness interchangeably.}.  
In the literature, \textbf{Provider-side Utility}\footnote{we use item fairness and provider-side fairness interchangeably} is used to measure a ranking system's ability to create a fair environment for items and their providers.
Since a fair environment should let similar items be treated similarly, items of similar relevance should get similar exposure in this fair ranking system, i.e., the amortized fairness principle~\cite{singh2018fairness,biega2018equity}.
Following existing works~\cite{singh2018fairness,biega2018equity,yang2021maximizing,singh2019policy}, item $d$'s exposure is defined as its accumulated examination probability:
\begin{equation}
\vspace{0pt}
\begin{split}
    E(d)&= \sum_{i=1}^t \sum_{j=1}^k p_{j}\mathbbm{1}_{\pi_i[j]==d} 
    \end{split}
    \label{eq:expo}
\vspace{0pt}
\end{equation}
where $\pi_i[j]$ indicates the $j^{th}$ item in ranked list $\pi_{i}$, $\mathbbm{1}$ is the indicator function which indicates that $p_{j}$ will contribute $E(d)$ only when $\pi_i[j]$ is item $d$.
% \brutuscomment{not sure what this means}
% Specifically, we can quantify the (un)fairness as the distance between relevance and exposure\cite{biega2018equity}.
% \begin{equation}
%     \textit{unfairness}(q)=\textit{Distance}(R,E)= \int_{d\in D(q)} \textit{Distance}(R(d), E(d))
%     \label{eq:unfairnessMetrics}
% \end{equation}
With items' exposure, we follow \cite{oosterhuis2021computationally} to define the unfairness as:
\begin{equation}
\begin{split}
    \textit{unfairness} &= \frac{1}{n(n-1)}\sum_{d_x\in D(q)} \sum_{d_y\in D(q)}\bigg(E(d_x)R(d_y)-E(d_y)R(d_x)\bigg)^2\\
    \label{eq:unfairness}
    \textit{fairness} &= -\textit{unfairness}
    % \label{eq:fairness}
\raisetag{2\normalbaselineskip}
\vspace{0pt}
\end{split}
\end{equation}
where $D(q)$  is the set of candidate items for query $q$. 
This unfairness measures the average exposure disparity between item pairs, and the fairness is just the negative of unfairness.
% For simplicity, we only give formulation where there is only one single query in this paper. 
% For the case of multiple queries, we just use the average value. 
Besides, we use $R$ or $R(d)$ as $R(d,q)$ for simpler notation in later formulation.
% Then we define fairness as
% \brutuscomment{use one sentence to explain this definition or reference to previous works}

% Item fairness. As we discussed in the introduction, sorting based ranking may provide optimal utility to users, but the ranking can be unfair, as few items get the majority of exposure. To quantify such fairness, we formally define fairness metric in this section. Firstly, we define the exposure as probability of examination. Since an item can be shown to users multiple times at different time steps, the exposure of an item is the accumulation of exposure it gets in history. Then, we adopt the amortizing fairness definition where an item’s exposure should be proportional to its relevance. 
\subsection{Partial and Biased Feedback}
\label{sec:biasedFeedback}
As shown in the ranking service's workflow in Figure~\ref{fig:workflow}, we rely on users' feedback to update the relevance estimation.
However, users' feedback could be partial and biased indicator of relevance since users only provide meaningful feedback for items that they have examined, we have:
\begin{equation}
\vspace{0pt}
    c = \begin{cases}
      r, & \text{if}\quad e=1 \\
      0, & \textit{otherwise}
    \end{cases}
\vspace{0pt}
\end{equation}
where binary random variable $e$ indicates whether an item has been examined by the user or not; binary random variable $r$ indicates whether an item is perceived as relevant by the user; binary random variable $c$ indicates whether an item is clicked\footnote{in this paper, we use feedback and click interchangeably.} by the user. 
Following existing works on click model~\cite{chuklin2015click,yang2021maximizing}, we model the probability of click $c$ as:
\begin{equation}
\vspace{0pt}
    p(c=1)=p(r=1)p(e=1)
    \label{eq:clickProb}
\vspace{0pt}
\end{equation}
where $p(e=1)$ is users' examination probability.  
Following existing works\cite{oosterhuis2021unifying,yang2022CanClicks}, we assume two kinds of biases exist in examination probability.

\noindent
\textit{\textbf{Positional Bias}}~\cite{craswell2008experimental,joachims2017accurately}: The  examination probability drops along ranks (also called position), and we model it with $p(rank(d|\pi))$, where the examining probability only depends on the rank.

\noindent
\textit{\textbf{Selection Bias}} \cite{oosterhuis2021unifying,oosterhuis2020policy}: This bias exists when not all of the items are selected to be shown to users or some lists are so long that no user will examine the entire lists. 
We model this by assuming that items ranked lower than rank $k_s$ won’t be examined by the user:
\begin{equation}
\vspace{0pt}
    P(e=1|d,\pi)=\begin{cases}
      p(\textit{rank}(d|\pi)), & \text{if}\quad \textit{rank}(d|\pi)\le k_s \\
      0, & \textit{otherwise}
    \end{cases}
    \label{eq:OverallBias}
% \vspace{0pt}
\end{equation}

\section{PROPOSED METHOD}
The challenge of optimizing fairness and effectiveness lies in the fact that the ranking optimization is being carried out based on relevance estimation while relevance estimation is still being learned.
% based on users' partial and biased feedback, i.e., in the online scenario. 
From a statistical point of view, an estimation such as relevance estimation mostly contains uncertainty, i.e., variance, and possibly some bias, which can make the optimization of fairness and effectiveness suboptimal.
For example, an item with under-estimated relevance and high uncertainty might never get presented to users since both optimizing fairness and optimizing effectiveness will rank irrelevant items to the lower positions of the ranked lists. No presentation will make this item hard to collect feedback to effectively update its relevance estimation.
However, this item might be actually relevant and we will know its relevance if more user feedbacks are collected to reduce the uncertainty in its relevance estimation.
Although many existing methods can give an unbiased estimation of relevance~\cite{oosterhuis2021unifying,yang2022CanClicks}, the uncertainty (i.e., variance) in this estimation might still make the relevance estimation unreliable, which will make them deliver suboptimal ranking results. 
To address uncertainty, we propose a Marginal-Certainty-aware Fair ranking algorithm called MCFair, which jointly optimizes effectiveness and fairness. 
% in an online setting. 
% \brutuscomment{inconsistent with the previous name}

For the rest of this section,
we first propose a gradient-based ranking optimization framework to optimize fairness and effectiveness where we assume uncertainty  does not exist (\S \ref{sec:gradientBased}); 
then we further extend this framework to be uncertainty-aware (\S\ref{sec:marginal-uncer-framework}).
Finally, we introduce the specific relevance estimator used for the proposed framework (\S\ref{sec:relEsti}).

\subsection{Gradient-based Optimization Framework}
\label{sec:gradientBased}
In this section, we introduce a gradient-based ranking optimization framework. 
As shown in Figure~\ref{fig:workflow}, at time step $t$, one user issues a query $q$ and the objective of the framework is to find the optimum ranked list $\pi_t$ that jointly maximizes effectiveness and fairness:
\begin{equation}
    \max_{\pi_t} \quad \textit{Obj.}(q,t)= \textit{eff.}(q,t)+ \alpha \textit{fair.}(q,t)
    \label{eq:eff-fair-joint}
\end{equation}
where $\alpha$ is the coefficient to balance the two utility. 
In this ranking objective, according to Eq. \ref{eq:NDCG} and Eq. \ref{eq:cumDCG}, we can reorganize the effectiveness as:
\begin{equation}
\vspace{0pt}
\textit{eff.}(q,t) = \textit{cNDCG}@k_s=\frac{1}{\textit{DCG}@k_s(\pi^*)}\sum_{\tau=1}^t\gamma^{t-\tau} \textit{DCG}@k_s(\pi_\tau)
\label{eq:effReorg}
\raisetag{2\normalbaselineskip}
\vspace{0pt}
\end{equation}  
In this paper, we will adopt $k_s$ (defined in Eq. \ref{eq:OverallBias}) as the cutoff for ranking effectiveness optimization unless explicitly specified.
Besides the cutoff, we will set $\gamma=1$ in Eq. \ref{eq:effReorg} for simplicity, and we will relax it to all $\gamma$ within $[0,1]$ in later discussion.  By ignoring the constant  $\textit{DCG}@k_s(\pi^*)$, we can get the $\textit{eff.}$ as:
% $$
% \textit{eff.}(q,t) = \textit{cNDCG}@k_s = \sum_{\tau=1}^t \textit{DCG}@k_s(\pi_\tau)=\sum_{i=1}^t \sum_{j=1}^k p_{j} R(\pi_i[j])\\
% $$
\begin{equation}
\begin{split}
\textit{eff.}(q,t) &= \textit{cNDCG}@k_s = \sum_{\tau=1}^t \textit{DCG}@k_s(\pi_\tau)=\sum_{i=1}^t \sum_{j=1}^k p_{j} R(\pi_i[j])\\
&=\sum_{d\in D(q)}\big( \smash[lr]{\sum_{i=1}^t \sum_{j=1}^k p_{j}R(d)\mathbbm{1}_{\pi_i[j]==d}\big)} \\&= \sum_{d\in D(q)}R(d)\big(\sum_{i=1}^t \sum_{j=1}^k p_{j}\mathbbm{1}_{\pi_i[j]==d}\big) \\
&= \sum_{d\in D(q)}R(d) E(d)
\raisetag{2\normalbaselineskip}
\label{eq:NDCGwithExpo}
\end{split}
\end{equation}
where $p_{j}$ is the examination probability of $j^{th}$ rank and $\pi_i[j]$ indicates the $j^{th}$ item in user $i$'s ranked list $\pi_{i}$. 
$E(d)$ is the cumulative exposure defined Eq. \ref{eq:expo}.
The above effectiveness formulation means that we should allocate more exposure $E(d)$ to items with greater relevance $R(d)$.

At time step $t$, the optimization goal defined in Eq.~\ref{eq:eff-fair-joint} is actually equivalent to finding $\pi_t$ to maximize the marginal objective, denoted as $\Delta \textit{Obj.}(q,t)$:
\begin{equation}
\begin{split}
       \max_{\pi_t} \;\textit{Obj.}(q,t)
       &\equiv \max_{\pi_t} \quad \textit{Obj.}(q,t) -\textit{Obj.}(q,t-1) \\
       &=\max_{\pi_t} \quad \Delta \textit{Obj.}(q,t)\\
       &=\max_{\pi_t} \quad \Delta \textit{eff.}(q,t)+\alpha \Delta \textit{fair.}(q,t)
\end{split}
\raisetag{2\normalbaselineskip}
\label{eq:marginalObjTransf}
\end{equation}
where the marginal objective is the increment of objective at time step $t$. $\equiv$ means equivalence.
The equivalence is due to the fact that the ranked list $\pi_t$ at time step $t$ won't change $\textit{Obj.}(t-1)$ because $\pi_t$ cannot change history.

To optimize $\Delta \textit{Obj.}(q,t)$, we take the first order approximation of the above marginal objective $\Delta \textit{Obj.}(q,t)$ by considering marginal exposure $\Delta E$'s influence:
\begin{equation}
\begin{split}
      \quad  &\Delta \textit{Obj.}(q,t)\approx \sum_{d\in D(q)}\frac{\partial{\textit{eff.}}}{\partial E(d)}\Delta E(d)+\alpha\sum_{d\in D(q)}\frac{\partial{\textit{fair.}}}{\partial E(d)}\Delta E(d)  \\
      &= \sum_{d\in D(q)} R(d) \Delta E(d)\\+&\alpha \sum_{d\in D(q)} \underbrace{ 
      \frac{4}{n(n-1)}\bigg(R(d)\sum_lE(l)R(l)-E(d)\sum_hR^2(h)\bigg)
      }_\text{B(d)} \Delta E(d) \\
      &=\sum_{d\in D(q)} \underbrace{ (R(d)+\alpha B(d))}_{g(d)} \Delta E(d)
% &=\sum_{d\in D(q)} g(d)\Delta E(d)\\
      =\langle \vec{g}\;,\; \Delta \vec{E}\rangle
\end{split}
\raisetag{2\normalbaselineskip}
\label{eq:MarginalObjective}
\end{equation}
where $\vec{g}$ is  $[g(d) \: \forall d \in D(q)]$, the vector form of gradients, $\langle,\rangle$ denotes dot product.
The marginal objective at time step $t$ can be approximated by the dot product of gradient $\vec{g}$ and marginal exposure $\Delta \vec{E}$ at time step $t$, i.e. $\langle \vec{g}\;,\; \Delta \vec{E}\rangle$.  
Actually, the gradient of effectiveness is $R(d)$, regardless of whether $\gamma=1$ or $0\le\gamma\le1$, since $\gamma$ only affect how we weight the historical DCGs (see Eq.~\ref{eq:effReorg}) in effectiveness and historical DCGs will not affect the current DCG at time step $t$, i.e., the marginal effectiveness. 
So, the above derivation still holds when  $0\le\gamma\le1$. The marginal exposure $\Delta \vec{E}$ is the exposure each item will get at time step $t$, i.e., $p_{j}$ in Eq. \ref{eq:NDCGwithExpo}. 
Since $p_{j}\in (0,1)$ is relatively small, the objective's first-order approximation should approximate $\Delta \textit{Obj.}(q,t)$ well.  
Furthermore, we can reformulate $\Delta \textit{Obj.}(q,t)$ as: 
\begin{equation}
\begin{split}
      \quad  \Delta \textit{Obj.}(q,t)&\approx 
      \sum_{d\in D(q)} g(d) \Delta E(d) =\sum_{k=1}^{|D(q)|}g(\pi_t[k])p_{k}
\end{split}
\end{equation}
where $\pi_t[k]$ is the $k^{th}$ item in the ranklist $\pi_t$.
To maximize the above $\Delta \textit{Obj.}(q,t)$, we first introduce the Rearrangement Inequality (refer to Section 10.2, Theorem 368 in \cite{hardy1952inequalities}). 
\begin{lemma}
Given $0\le x_1\le x_2 \le ... \le x_n$ and $y_1\le y_2 \le ... \le y_n$, we have:
\begin{equation}
    \sum_{i=1}^n x_{\sigma(i)}y_i \le \sum_{i=1}^n  x_iy_i
\vspace{0pt}
\end{equation}
where $\sigma$ can be any possible permutation. 
\end{lemma}
According to the above rearrangement inequality, we should let item with greater gradient $g(d)$ get greater examination probability $p_k$ in order to maximize $\Delta \textit{Obj.}(q,t)$. By assuming that $p_{k}$ drops as rank $k$ increases, \textbf{we can maximize $\Delta \textit{Obj.}(q,t)$ by generating a ranked list $\pi_t$ that arranges items according to their gradients $g$ in descending order}:
\begin{equation}
    \pi_t=\arg_{\textit{topk}}\big(g(d)| \forall d \in D(q)\big)
    \label{eq:sortRankingscore}
\end{equation}
where $k$ is the length of ranked list $\pi_t$.
% \subsubsection{Effectiveness vs. Fairness}
% \label{sec:fixedEff}

Aside from optimization, in this paper, we also reveal an interesting relation between effectiveness and fairness. 
When fairness constraint is strictly satisfied and unfairness is reduced to zero, $\textit{cNDCG}@k_s$ is actually fixed. 
Then, according to Eq. \ref{eq:unfairness}, we can reduce the unfairness to zero when items get exposure as:
\begin{equation}
\vspace{0pt}
    \bar{E}(d)=\frac{R(d)}{\sum_{h\in D} R(h)}E_{sum} \quad \forall d\in D(q)
\vspace{0pt}
\end{equation}
where the total exposure $E_{sum}=\sum_{d\in D} E(d)$. 
By setting ${E}(d)$ to $\bar{E}(d)$ in Eq. \ref{eq:NDCGwithExpo}, we can get the $\textit{cNDCG}@k_s$ as:
\begin{equation}
\vspace{0pt}
\begin{split}
   \overline{cNDCG}@k_s&=\sum_{d\in D(q)}R(d) E(d) = \frac{\sum_{d\in D(q)}R^2(d)}{\sum_{h\in D} R(h)}E_{\textit{sum}}
    \end{split}
\label{eq:fixedEff}
\vspace{0pt}
\end{equation}
where we still assume $\gamma=1$ and ignore the normalization. 
From the above derivation, we could know that $\textit{cNDCG}@k_s$ is fixed as long as the fairness constraint is strictly satisfied, no matter which algorithm we use to reach fairness. 
Although $\textit{cNDCG}@k_s$ is fixed, it still leaves us much freedom to improve the top ranks' effectiveness $\textit{cNDCG}@k_c$ ($k_c<k_s$) as well as to improve effectiveness when some tolerance of fairness is allowed.

\subsection{Uncertainty-aware Ranking Optimization}
\label{sec:marginal-uncer-framework}
In this section, we will extend the above gradient-based ranking optimization framework to be uncertainty-aware. 

In the above ranking optimization,  we optimize ranking in  \textbf{a post-processing setting} where relevance $R$ is already well-estimated prior to ranking optimization, and we can calculate $g$ in Eq. \ref{eq:MarginalObjective} as the ranking score based on the pre-estimated relevance.  
Such an assumption means we optimize ranking in a post-processing manner. 
However, in real-world applications, relevance estimation and ranking optimization are often entangled in \textbf{an online setting}, where ranking optimization takes place while relevance estimation is still being learned.
% The problem brought by the online setting is that relevance estimation is often not perfect and contains uncertainty (variance).  %Zhenduo
The online setting brings us a problem relevance estimation is often not perfect and contains uncertainty (variance).  
% In this paper, we assume that the well estimated relevance, denoted as $R(d)$, in the post-processing setting is trustworthy and we only consider the uncertainty in the online setting.
In this section, we focus on the scenario where relevance is estimated online with uncertainty.

To analyze uncertainty, we first accumulate the uncertainty of all candidate items:
\begin{equation}
    \reallywidehat{\textit{uncert.}}(q,t)=\sum_{d\in D(q)} \textit{Variance}(\hat{R}(d))
\end{equation}
where $\hat{R}(d)$ is item $d$'s relevance estimation\footnote{Notation with a $\hat{}$
is used to denote something is an estimation and contains uncertainty.} and $\textit{Variance}(\hat{R}(d))$ is the variance of $\hat{R}(d)$; we leave co-variance to future work.
As for how to estimate relevance, we will discuss it in \S\ref{sec:relEsti}. 
% For simplicity, we leave co-variance for future discussion.  
Being uncertainty-aware, we formulate the ranking objective as:
\begin{equation}
      \max_{\pi_t} \quad \textit{Obj.}(q,t)= \reallywidehat{\textit{eff.}}(q,t)+ \alpha\reallywidehat{\textit{fair.}}(q,t)-\beta \reallywidehat{\textit{uncert.}}(q,t)
      \label{eq:maginalObjWithUncer}
\end{equation}
where $\reallywidehat{\textit{eff.}}$ and $\reallywidehat{\textit{fair.}}$ are the estimated effectiveness and fairness, calculated by substituting $R$ with $\hat{R}$ in Eq. \ref{eq:NDCGwithExpo} and Eq. \ref{eq:unfairness}. 
In this paper, we use the hatted notation, when it is calculated based on relevance estimation $\hat{R}$.  

Although our goal is to jointly maximize effectiveness and fairness, we still include the negative $\reallywidehat{\textit{uncert.}}$ as a third goal to decrease the uncertainty when optimizing $\pi_t$. 
We follow the assumption that decreasing uncertainty to get a more certain relevance estimation for candidate items can help better optimize effectiveness and fairness, which is later verified by our experimental results. 
To optimize the above ranking objective, we follow Eq. \ref{eq:marginalObjTransf} and Eq. \ref{eq:MarginalObjective} to take the first order approximation of the marginal objective by considering marginal exposure $\Delta E(d)$,
% \begin{equation}
% \begin{split}
%       \max \quad  \Delta \textit{Obj.}&\approx \sum_i \underbrace{R_i}_\text{part 1} \Delta E_i+\alpha \sum_i \underbrace{B_i}_\text{part 2} \Delta E_i-\beta \sum_i \underbrace{\frac{\partial\textit{uncert.}}{\partial E_i}} _\text{part 3}\Delta E_i \\
%       &=\sum_i (R_i+\alpha B_i+\beta \frac{1}{E_i^2}) \Delta E_i 
% \\
%       &=\sum_i g_i\Delta E_i\\
%       &=\langle \vec{g}\;,\; \Delta \vec{E}\rangle
% \end{split}
% \label{eq:MarginalObjective}
% \end{equation}
\begin{equation}
\begin{split}
    \quad  \Delta \textit{Obj.}(t)
    &\approx \sum_{d\in D(q)}\bigg(\frac{\partial{\reallywidehat{\textit{eff.}}}}{\partial E(d)}+\alpha\frac{\partial{\reallywidehat{\textit{fair.}}}}{\partial E(d)}+
    \beta \underbrace{\frac{-\partial{\reallywidehat{\textit{uncert.}}}}
    {\partial E(d)}}_\textit{\reallywidehat{MC}}\bigg)\Delta E(d) \\
    &=\sum_{d\in D(q)} \underbrace{ \bigg(\hat{R}(d)+\alpha\hat{B}(d)+\beta \reallywidehat{MC}(d)\bigg)}_{\reallywidehat{ug}(d)} \Delta E(d) \\
    %   &=\sum_{d\in D(q)} \reallywidehat{ug}(d)\Delta E(d)\\
      &=\langle \vec{\reallywidehat{ug}}\;,\; \Delta \vec{E}\rangle
\end{split}
\raisetag{2\normalbaselineskip}
\label{eq:MarginalObjectiveUnc}
\end{equation}
% \begin{equation}
% \begin{split}
%       \quad  \Delta \textit{Obj.}(t)
%       &\approx \sum_{d\in D(q)}\bid(\frac{\partial{\reallywidehat{\textit{eff.}}}}{\partial E(d)}+\alpha\frac{\partial{\reallywidehat{\textit{fair.}}}}{\partial E(d)}+\beta\underbrace{ \frac{-\partial{\reallywidehat{\textit{uncert.}}}}{\partial E(d)}}_\textit{Marginal certainty}\big)\Delta E(d) \\
%       &=\sum_{d\in D(q)} \underbrace{ (\reallywidehat{g}(d)+\beta\frac{1}{E(d)^2})}_{\reallywidehat{ug}(d)} \Delta E(d) 
% \\
%     %   &=\sum_{d\in D(q)} \reallywidehat{ug}(d)\Delta E(d)\\
%       &=\langle \vec{\reallywidehat{ug}}\;,\; \Delta \vec{E}\rangle
% \end{split}
% \label{eq:MarginalObjectiveUnc}
% \end{equation}
where $\reallywidehat{MC}$ denotes Marginal Certainty.
Recall that in Eq.~\ref{eq:sortRankingscore}, directly using $\reallywidehat{ug}(d)$ as ranking scores to generate the ranklist $\pi_t$ can help optimize the ranking objective.
% \begin{equation}
%     \pi_t=\arg_{topk}\big(g(d)| \forall d \in D(q)\big)
%     \label{eq:sortRankingscoreUn}
% \end{equation}
% where $k$ is the length of ranked list $\pi_t$.
By optimizing such ranking objective, we automatically get a marginal-certainty-aware exploration strategy. 
With this strategy, items bringing greater marginal certainty $\reallywidehat{MC}$ will be boosted in $\reallywidehat{ug}$ to increase their exposure. 
We refer to the marginal certainty based fairness optimization method as \textbf{MCFair}. 
We also notice a recent related work UCBRank~\cite{yang2022CanClicks}, which directly boosts items' ranking scores with uncertainty for exploration. 
Our marginal-certainty-aware exploration strategy is different from UCBRank, as we consider marginal (un)certainty instead of the uncertainty itself. 
And we believe that marginal (un)certainty is more effective in terms of exploration. 
For example, if there exit some items of high uncertainty and such uncertainty cannot be reduced with more user interaction, i.e., low marginal certainty, we shouldn't boost their scores because boosting them cannot reduce the uncertainty of the relevance estimation. 
Therefore we adopt marginal (un)certainty because we think it could better guide the exploration than the uncertainty itself.
% In this paper, we limit our discussion within the scope of fair ranking algorithms and UCBRank is not a ranking algorithm to address fairness. We leave comparison between UCBRank in future works.
% So it is marginal (un)certainty instead of uncertainty itself that should guide the exploration.

\subsection{Unbiased Relevance Estimator}
\label{sec:relEsti}
% In the above gradient based ranking optimization framework, any relevance estimator can be included. 
The aforementioned gradient-based ranking optimization framework  does not depend on the specific choice of relevance estimator.
Here we introduce unbiased relevance estimator we adopt in this work. 
As user feedback could be biased relevance indicator, we follow previous works \cite{yang2022CanClicks,oosterhuis2021unifying} and use a unbiased estimator of the relevance:
\begin{equation}
\vspace{0pt}
    \hat{R}(d)=\frac{\textit{cumC}(d)}{E(d)}
    \label{eq:rele-estimation}
\vspace{0pt}
\end{equation}
where $\textit{cumC}(d)$ is the cumulative clicks computed by:
\begin{equation}
\begin{split}
    \textit{cumC}(d)&= \sum_{i=1}^t \sum_{j=1}^k c_{i,j}\mathbbm{1}_{\pi_i[j]==d} 
    \end{split}
    \label{eq:clicksum}
\end{equation}
The $\hat{R}(d)$ is an unbiased estimation of the relevance $R(d)$:
\begin{equation}
 \begin{split}
   \E_c[\hat{R}(d)]&= \E_c[\frac{\textit{cumC}(d)}{E(d)}]
   =\frac{\sum_{i=1}^t \sum_{j=1}^k \E_c[c_{i,j}]\mathbbm{1}_{\pi_i[j]==d} }{\sum_{i=1}^t \sum_{j=1}^k p_{j}\mathbbm{1}_{\pi_i[j]==d} }\\
   &=\frac{\sum_{i=1}^t \sum_{j=1}^k p_{j}\times p(R=1|d,q)\mathbbm{1}_{\pi_i[j]==d} }{\sum_{i=1}^t \sum_{j=1}^k p_{j}\mathbbm{1}_{\pi_i[j]==d} }\\   
   &=p(R=1|d,q) \times \frac{\sum_{i=1}^t \sum_{j=1}^k p_{j}\mathbbm{1}_{\pi_i[j]==d} }{\sum_{i=1}^t \sum_{j=1}^k p_{j}\mathbbm{1}_{\pi_i[j]==d}}\\  
   &=p(R=1|d,q) =R(d)
\end{split}
\end{equation}
And we can also compute $\hat{R}(d)$'s variance by:
\begin{subequations}
\begin{align}
    \textit{Variance}[\hat{R}(d)]&=\frac{\sum_{i=1}^t \sum_{j=1}^k \textit{Var}[c_{i,j}]\mathbbm{1}_{\pi_i[j]==d}}{E(d)^2} \label{eq:combination}\\
    &=\frac{\sum_{i=1}^t \sum_{j=1}^k (\E_c[c_{i,j}^2]-\E_c^2[c_{i,j}])\mathbbm{1}_{\pi_i[j]==d}}{E(d)^2}\\
    % &=\frac{\sum_{i=1}^t \sum_{j=1}^k (\E_c[c_{i,j}]-\E_c^2[c_{i,j}])\mathbbm{1}_{\pi_i[j]==d}}{E(d)^2}\label{eq:Cbinary}\\    
    &=\frac{\sum_{i=1}^t \sum_{j=1}^k (p(c_{i,j}=1)-p(c_{i,j}=1)^2)\mathbbm{1}_{\pi_i[j]==d}}{E(d)^2}\label{eq:CbinaryExptoprob}\\   
    &=\frac{\sum_{i=1}^t \sum_{j=1}^k (p_{j}R-p_{j}^2R^2)\mathbbm{1}_{\pi_i[j]==d}}{E(d)^2}\\
    &=\frac{\sum_{i=1}^t \sum_{j=1}^k (p_{j}R-p_{j}^2R^2)\mathbbm{1}_{\pi_i[j]==d}}{E(d)^2}\\
    &< \frac{\sum_{i=1}^t \sum_{j=1}^k p_{j}R\mathbbm{1}_{\pi_i[j]==d}}{E(d)^2} = \frac{R E(d)}{E(d)^2}
    < \frac{1}{E(d)}
    \raisetag{2\normalbaselineskip}
    \vspace{-5pt}
\end{align}  
\end{subequations}
For simplicity, we use above upper bound as the variance. 
In Eq.~\ref{eq:combination}, we treat $\hat{R}(d)$ as a linear combination of $c_{i,j}$ to get the variance. 
In Eq.~\ref{eq:CbinaryExptoprob}, $c_{i,j}=c_{i,j}^2$ and $\E_c[c_{i,j}]=p(c_{i,j}=1)$ since $c_{i,j}$ is binary random variable. 
% In Eq. ~\ref{eq:CbinaryExptoprob}, we have
% $\E_c[c_{i,j}]=p(c_{i,j}=1)$.
% \begin{equation}
% \E_c[c_{i,j}]=p(c_{i,j}=1)\times 1+p(c_{i,j}=0)\times 0=p(c_{i,j}=1)
% \end{equation}
With above variance, we could get the $\reallywidehat{MC}(d)$ as:
\begin{equation}
    \reallywidehat{MC}(d)=\frac{1}{E(d)^2}
    \label{eq:MC}
\end{equation}
% we can get the final ranking score $\reallywidehat{ug}(d)$ as 
% \begin{equation}
%     \reallywidehat{ug}(d)=\reallywidehat{R}(d)+\alpha\reallywidehat{B}(d)+\beta \frac{1}{E(d)^2}
% \end{equation}
By substituting $\reallywidehat{MC}(d)$ to Eq. ~\ref{eq:MarginalObjectiveUnc}, we will get ranking score $\reallywidehat{ug}(d)$. The ranked list is generated by $\pi_t=\arg_{topk}\big(\reallywidehat{ug}(d) | \forall d\in D(q)\big)$. 
% \begin{equation}
%     \pi_t=\arg_{topk}\big(\reallywidehat{ug}(d) | \forall d\in D(q)\big)
%     \label{eq:sortRankingscoreUn}
% \end{equation}
In this paper, we only use non-parameterized relevance estimator in Eq. \ref{eq:rele-estimation}. 
% For parameterized relevance estimator based on features, some complicated methods have been proposed to compute its marginal uncertainty. \brutuscomment{add citation}
Due to the space limit, we leave the analysis of parameterized relevance estimator and its marginal uncertainty to future work.

\section{EXPERIMENTS}
\subsection{Experimental Setup}
To evaluate our methods, we will conduct semi-synthetic experiments. 
We cover the experimental settings in this section. 
% All the experimental scripts and implementations used will be available online\footnote{\url{http://github.com/hide-for-anonymous-review}}.
\subsubsection{Datasets}. In the experiment, we use two publicly available datasets, i.e., MQ2008~\cite{qin2013introducing} and Istella-S~\cite{lucchese2016post}. MQ2008 has three-level relevance judgments (from 0 to 2), and  Istella-S has five-level relevance judgments (from 0 to 4). MQ2008 has about 800 queries and about 20 candidate documents for each query. Istella-S has about 33018 queries and about 103 candidate documents for each query. Queries in both datasets are divided into training, validation, and test partitions. 
\subsubsection{Baselines}.
\label{sec:baselines}
To evaluate the proposed method, we compare the following baselines,
\begin{itemize}[leftmargin=*]
    \vspace{-3pt}
    \item \textbf{TopK}. Sort items according to  relevance $\reallywidehat{R}(d)$, i.e., the first part of $\reallywidehat{ug}(d)$ in Eq.~\ref{eq:MarginalObjectiveUnc}.
    \item \textbf{FairK}. Sort items according to the gradient of fairness $\reallywidehat{B}(d)$, i.e., the second part of $\reallywidehat{ug}(d)$ in Eq.~\ref{eq:MarginalObjectiveUnc}.
    \item \textbf{ExploreK}. Sort items according to marginal certainty $\reallywidehat{MC}(d)$, i.e., the third part of $\reallywidehat{ug}(d)$ in Eq.~\ref{eq:MarginalObjectiveUnc}.
    \item \textbf{FairCo}~\cite{morik2020controlling}. Fair ranking  algorithm based on a proportional controller. $\alpha \in [0.0,1000.0]$
    \item \textbf{ILP}~\cite{biega2018equity}. Fair ranking  algorithm based on Integer Linear Programming (ILP).$\alpha \in [0.0,1.0]$
    \item \textbf{LP}~\cite{singh2018fairness}. Fair ranking  algorithm based on Linear Programming (LP).$\alpha \in [0.0,1000.0]$
    \item \textbf{MMF}~\cite{yang2021maximizing}. Similar to FairCo but focus on top ranks fairness. $\alpha \in [0.0,1.0]$
    \item \textbf{PLFair}~\cite{oosterhuis2021computationally}. Fair ranking  algorithm based on Placket-Luce optimization. $\alpha \in [0.0,1.0]$
    \item \textbf{MCFair}. Our method. Sort items according to the gradient of fairness $\reallywidehat{ug}(d)$ in Eq.~\ref{eq:MarginalObjectiveUnc}. $\alpha \in [0.0,1000.0]$
    \vspace{0pt}
\end{itemize}
Among the above ranking algorithm, TopK and ExploreK are unfair algorithms, while the others are fair algorithms. 
Among the fair algorithms, FairK directly uses fairness's gradient to rank items and can be viewed as a reduced and degenerated  version of MCFair when MCFair's $\alpha$ is set to a large number. Please note that FairK is also our proposed method which is derived with the gradient-based optimization framework proposed in Sec~\ref{sec:gradientBased}.
Except FairK, all other fair ranking algorithms have trade-off parameters to balance effectiveness and fairness, referred to as $\alpha$. 
Given a greater tradeoff parameter $\alpha$, the fair algorithms including FairCo, ILP, LP and MCFair care more about fairness, i.e., less tolerance for unfairness, which usually sacrificing effectiveness.
For example, MCFair can increase  $\alpha$ in Eq. \ref{eq:maginalObjWithUncer} to give a higher weight to fairness during optimization. 
For different fair algorithms, $\alpha$ may lie in different range.  
For FairCo, LP, and MCFair, $\alpha$ are within $[0.0,+\infty]$, and we adopt $\alpha\in [0.0,1000.0]$ in our experiment. 
For ILP, $\alpha\in [0.0,1.0]$. In this paper, we do not tune and select one $\alpha$ for each baseline when comparing each baseline. The reason is that  different ranking applications can have different fairness requirements, and one $\alpha$ for each baseline is not enough for covering different fairness requirements. To make a comprehensive comparison, we compare baselines for all $\alpha$ within each baseline's respective ranges instead of one particular tuned $\alpha$ within its range. Detailed comparison method can be found at Sec~\ref{sec:balancePost}.
% Since different ranking systems can have different fairness requirements, it is not enough to select one $\alpha$ (e.g., from validation) for each method and compare each method. Because of this, we mainly compare the fair algorithms within the whole $\alpha$ range instead of selecting one.
% Since different $\alpha$ (implicitly) mean different unfairness tolerance,
% In this paper, we mainly compare the fair algorithms within the whole $\alpha$ range instead of selecting one 
% do not seek to tune the optimal $\alpha$ for each baseline with validation in this paper. The reason is 
% For each of the fair algorithms, $\alpha$
% Please note that we do not seek to tune the best $\alpha$ for each baseline with validation in this paper. Instead we .
Besides, we limit our discussion within the scope of ranking fairness. UCBRank~\cite{yang2022CanClicks} is not chosen as a baseline since it does not address the ranking fairness problem.
% We leave comparison between UCBRank in future works.
% So it is marginal (un)certainty instead of uncertainty itself that should guide the exploration.

During implementation, we notice that methods LP and ILP are highly time-consuming as their decision variable quadratically increases with the number of candidates.  %Zhenduo 'really' is too colloquial, to 'highly'
Considering the time cost, we filtered out queries with more than 20 documents for MQ2008 and we didn't evaluate ILP and LP on the larger dataset, i.e., the Istella-S dataset.

\subsubsection{Ranking Service Simulation.} 
\label{sec:Simulation}
Following the workflow in Figure \ref{fig:workflow}, at each time step, a simulated user will issue a query by randomly sampling a query from the training, validation, or test partition. 
Then a ranking algorithm will construct a ranklist $\pi$ of candidate items and present it to the simulated user. To simulate user's click on the ranklist $\pi$, we need to simulate the relevance and examination (details in \S\ref{sec:biasedFeedback}). 
Following~\cite{ai2018unbiased}, the relevance probability of each document-query pair $(d,q)$ is simulated according to its relevance judgements $y$ as $P(r=1|d,q,\pi)=\epsilon+(1-\epsilon)\frac{2^y-1}{2^{y_{max}}-1}$
% \begin{equation}
%     P(r=1|d,q,\pi)=\epsilon+(1-\epsilon)\frac{2^y-1}{2^{y_{max}}-1}
% \end{equation}
where $y_{max}$ is the maximum value of relevance judgement $y$. $y_{max}$ can be 2 or 4 depending on the datasets. 
Aside from relevance, following~\cite{morik2020controlling,oosterhuis2021unifying}, we also simulate user's examination probability on $\pi$ as, $ P(e=1|d,\pi)=
      \frac{1}{\log_2(\textit{rank}(d|\pi)+1)}$.
% \begin{equation}
%     P(e=1|d,\pi)=\begin{cases}
%       \frac{1}{\log_2(\textit{rank}(d|\pi)+1)}, & \text{if}\quad \textit{rank}(d|\pi)\le k_s \\
%       0, & \text{otherwise}
%     \end{cases}
% \end{equation}
We only simulate users' examination behavior on top ranks, and we set $k_s$ to 5 throughout the experiments.  
For simplicity, we follow existing works~\cite{oosterhuis2021unifying,yang2022CanClicks,yang2021maximizing,morik2020controlling} to assume  that users' examination $P(e=1|d,\pi)$ is known in experiment as many existing works \cite{ai2018unbiased,wang2018position,agarwal2019estimating,radlinski2006minimally} have been proposed for it.
With simulated relevance and examination behavior, we sample and collect clicks with Eq. \ref{eq:clickProb}. 

Aside from the simulated users' behavior, we notice that LP and ILP methods were originally proposed with the assumption that relevance was already well-estimated prior to ranking optimization. 
However, in most real-world systems, ranking optimization and relevance learning are carried out at the same time.
To give a comprehensive comparison of different methods, we will compare two settings. 
The first setting is the \textbf{post-processing setting} where true relevance $R$ is already given.
The second one is the \textbf{online setting} where ranking optimization happens while relevance estimation is still being learned. 
In the post-processing setting, all the ranking methods in Section~\ref{sec:baselines} are based on true relevance $R$, and we assume true relevance $R$ is known in advance. 
Our method MCFair will set $\beta$ as 0. 
In the online setting, all the ranking methods in Section~\ref{sec:baselines} are based on the relevance estimation $\reallywidehat{R}$ in Eq. \ref{eq:rele-estimation} to perform ranking optimization, and we assume true relevance $R$ is not known at all. 
Our method MCFair will set $\beta$ to 100 unless otherwise explicitly specified, as 100 works well across all of our experiments.
For MQ2008, we simulate $10^4$ and  $10^5$ steps for post-processing and online settings, respectively. 
The online setting actually requires more iterations to learn relevance. 
For Istella-S, we simulate $10^6$ and $10^7$ steps for post-processing and online settings, respectively. 
% The experimental scripts and implementations used in this paper are available online\footnote{\url{https://github.com/Taosheng-ty/WSDM22-MCFair.git}}.

\subsubsection{Evaluation}. We use the cumulative NDCG (cNDCG) in Eq.~\ref{eq:cumDCG} with $\gamma=0.995$ to evaluate the effectiveness at different cutoffs, $1\le k_c\le 5$. 
Aside from effectiveness,  unfairness defined in Eq.~\ref{eq:unfairness} is used for unfairness measurement. 
We run each experiment five times and report the average evaluation performance on the test partition of each dataset. Note that the test partition was used only for evaluation and not for optimization or validation. Actually, as we mentioned in Sec~\ref{sec:baselines}, we do not tune and select parameters (e.g., $\alpha$) in this paper. We compare baselines in a comprehensive way where performance within its whole parameter space is considered.
Significant tests are conducted with the Fisher randomization test~\cite{smucker2007comparison} with $p<0.05$. Due to the time cost (see Table.~\ref{tab:performance}), we do not run ILP and LP on the larger Istella-S dataset, and the performances are not available. 
% To evaluate fair algorithms, FairCo, ILP, LP, and MCFair, we think it is not enough to compare  effectiveness and fairness with one single $\alpha$ selected according to validation since different ranking systems may have different effectiveness and fairness requirement. Instead, we choose to compare effectiveness-fairness joint performance within the whole range of $\alpha$, where we will know which one can get maximum effectiveness given the same fairness.

\subsection{Results in the Post-processing Setting.}
\begin{table}[t]
\label{tab:UnfTime}
\centering
    \caption{Unfairness and average time for generating 1k ranklists during simulation in the post-processing setting, where $\alpha$, if available, are set to the maximum value. Standard deviation is in the parentheses. By setting $\alpha$ to the maximum value, we compare algorithms' fairness capacity to mitigate unfairness. Time cost for ILP and LP on Istella-S are estimated by only run 1k steps instead of the total simulation steps indicated in Sec.~\ref{sec:Simulation}. Due to the time cost, unfairness performance  of ILP and LP  on the larger Istella-S dataset are NA in Table.}
    \centering
  \resizebox{1.0\columnwidth}{!}{
    \begin{tabular}{ccc|cc}\toprule
        Methods & \multicolumn{4}{c}{Datasets}\\ \cline{2-5}
        &{MQ2008} &{Istella-S} &{MQ2008} &{Istella-S}\\ 
        % & \multicolumn{4}{c}{Time(seconds)}\\ \cline{2-5}
        \hline
\textit{Unfair algorithm}&\multicolumn{2}{c|}{unfairness}&\multicolumn{2}{c}{Time (sec)}\\
        TopK&214.4$_{(3.884)}$	&19.45$_{(0.047)}$ &0.543$_{(0.159)}$&0.572$_{(0.131)}$ \\
        ExploreK&261.3$_{(7.128)}$	&3.452$_{(0.040)}$ &0.577$_{(0.192)}$&0.700$_{(0.053)}$\\ \hline
\textit{Fair algorithm} & & & &\\
        FairCo \cite{morik2020controlling} &23.69$_{(0.740)}$	&0.038$_{(0.005)}$&0.691$_{(0.176)}$&0.607$_{(0.015)}$\\ 
        LP \cite{biega2018equity} &25.44$_{(0.747)}$ &NA&4.036$_{(0.157)}$ &$\ge$10 days\\ 
        ILP \cite{singh2018fairness} &47.55$_{(1.718)}$ &NA&17.24$_{(0.479)}$&1508.9$_{(80.83)}$  \\
        MMF \cite{yang2021maximizing} &53.36$_{(1.334)}$ &0.154$_{(0.007)}$  &2.133$_{(0.205)}$ &6.876$_{(0.475)}$ \\
        PLFair \cite{oosterhuis2021computationally} & 256.4$_{(5.988)}$ &3.700$_{(0.062)}$&4.283$_{(0.309)}$ &4.366$_{(0.080)}$\\
        FairK(Ours)& 23.16$_{(0.742)}$	&0.030$_{(0.005)}$&0.627$_{(0.201)}$ &0.770$_{(0.099)}$\\
        MCFair(Ours)& \textbf{22.68}$_{(0.735)}$	&\textbf{0.029}$_{(0.005)}$&0.631$_{(0.195)}$&0.645$_{(0.011)}$ \\
        \bottomrule
        \end{tabular}}
    \label{tab:performance}
% \vspace{-5pt}
\end{table}
\begin{figure}[t]
    \centering
    \begin{subfigure}[b]{\columnwidth}
    \includegraphics[scale=0.45]{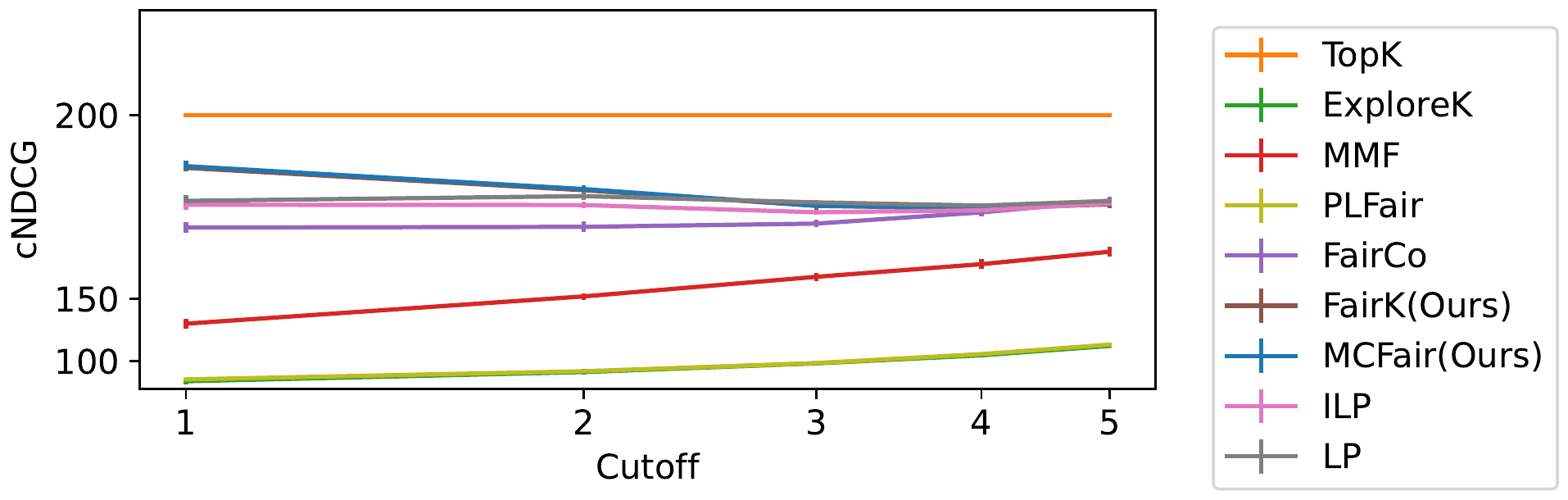}
    \vspace{-1.5\baselineskip}
    \caption{MQ2008 dataset}
    \label{fig:dist_a}
    \end{subfigure}
    \begin{subfigure}[b]{\columnwidth}
    \includegraphics[scale=0.45]{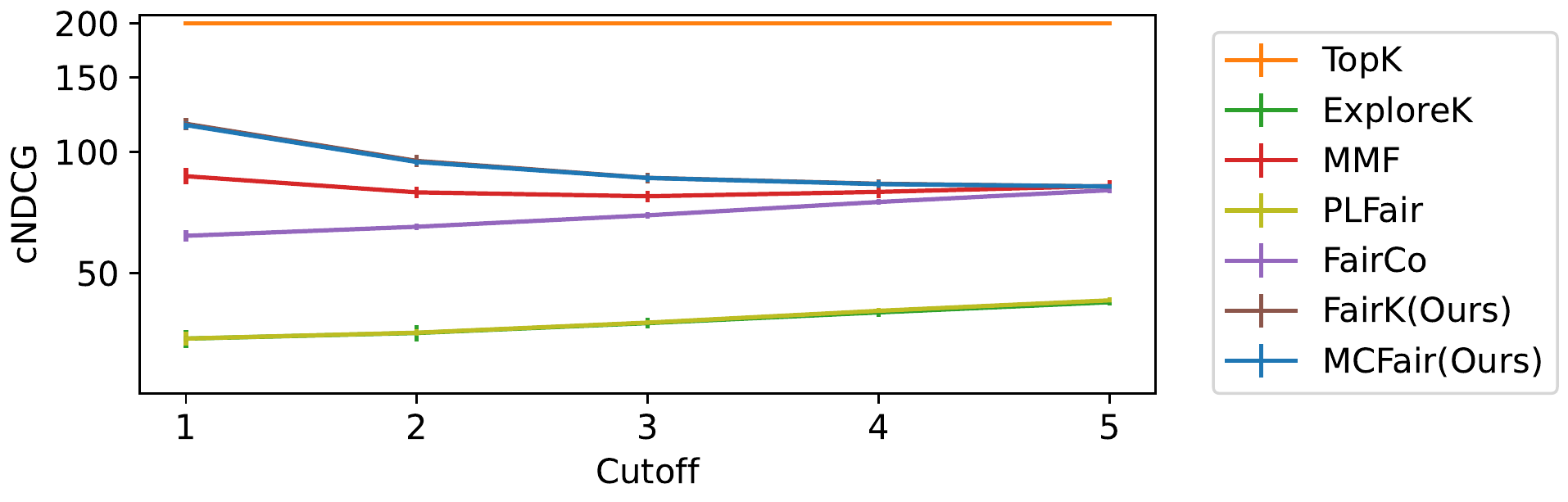}
    \vspace{-1.5\baselineskip}
    \caption{Istella-S dataset}
    \label{fig:dist_a}
    \end{subfigure}
    \caption{cNDCG of different prefix in the post-processing setting. FairK and MCFair overlap with each other, PLFair and ExploreK overlap.}
    \label{fig:NDCGcutoff}
% \vspace{-10pt}
\end{figure}
In this section, we first compare fair methods' fairness capacity, i.e., the maximum fairness a method can reach. 
Then we will discuss the effectiveness performance given different degree of fairness requirement.

\begin{figure*}[h]
    \vspace{-5pt}
    \centering
    \begin{subfigure}[]{0.22\textwidth}
    \includegraphics[scale=0.25]{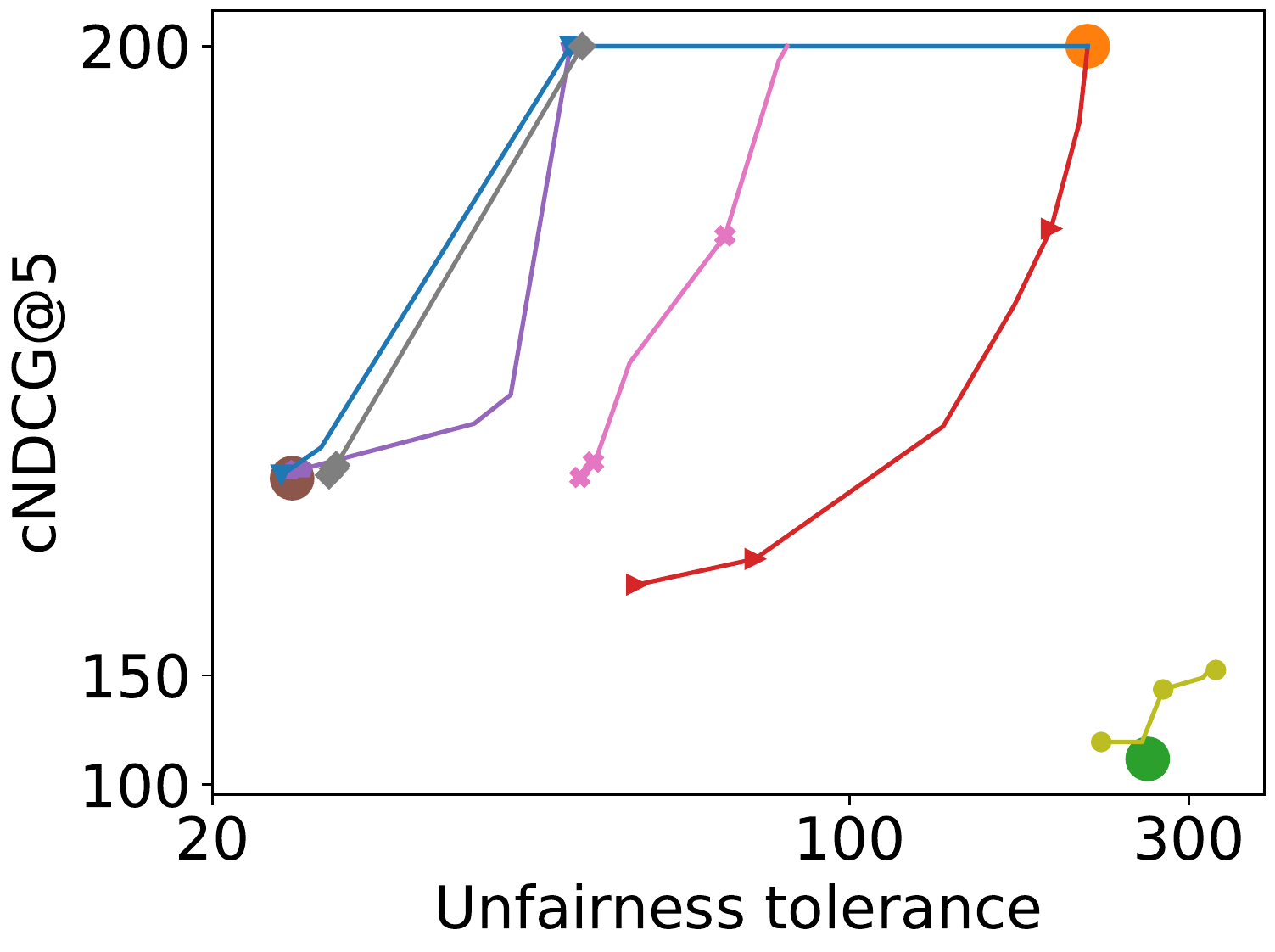}
    %\vspace{-0.5\baselineskip}
    \caption{MQ2008 (post-processing).}
    \label{fig:MQPost}
    \end{subfigure}
    \hfill
    \begin{subfigure}[]{0.22\textwidth}
    \includegraphics[scale=0.25]{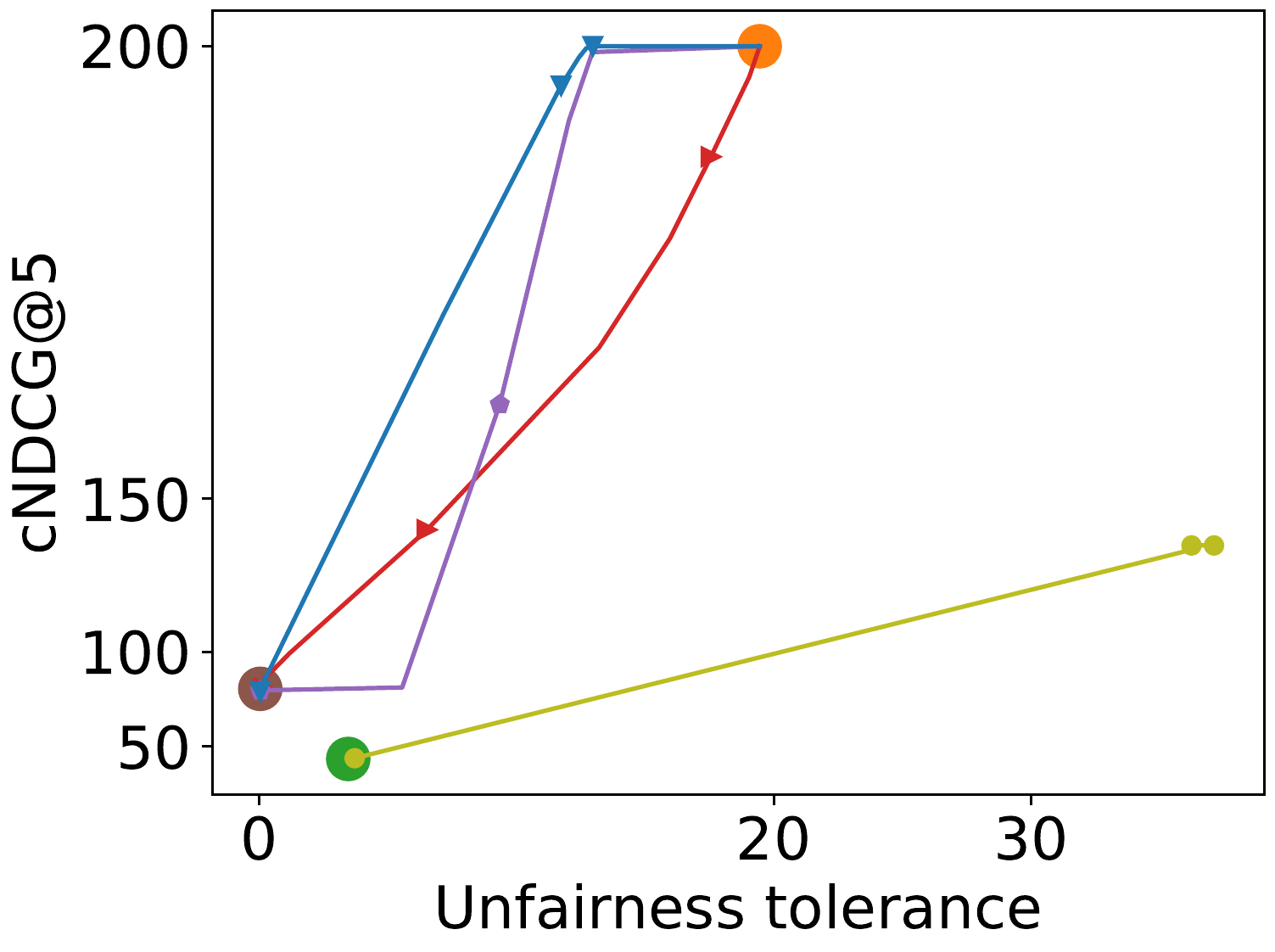}
    %\vspace{-0.5\baselineskip}
    \caption{Istella-S  (post-processing).}
    \label{fig:IstePost}
    \end{subfigure}\hfill
    \begin{subfigure}[]{0.22\textwidth}
    \includegraphics[scale=0.25]{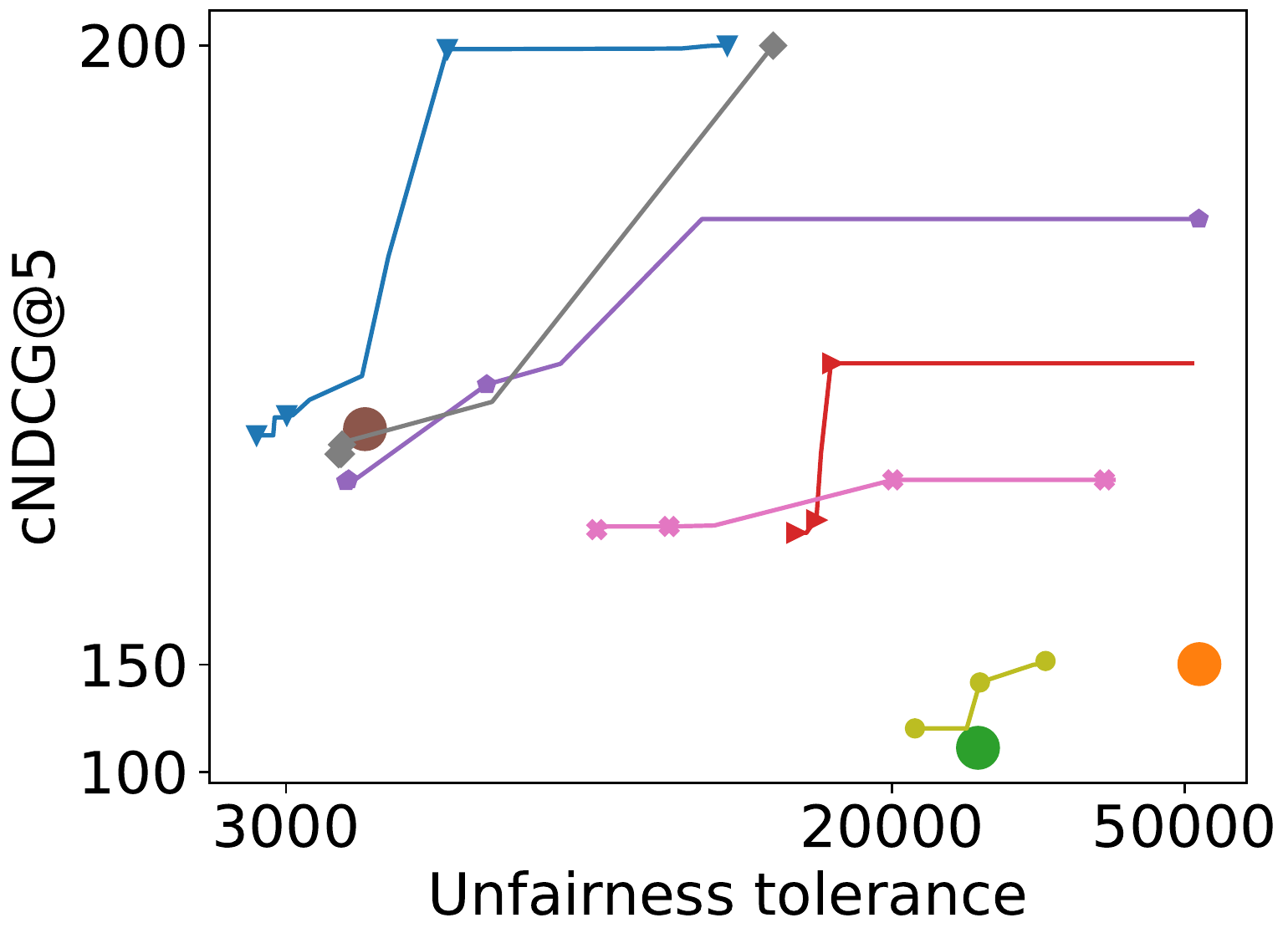}
    %\vspace{-0.5\baselineskip}
    \caption{MQ2008 (online).}
    \label{fig:MQOnline}
    \end{subfigure}\hfill
    \begin{subfigure}[]{0.22\textwidth}
    \includegraphics[scale=0.25]{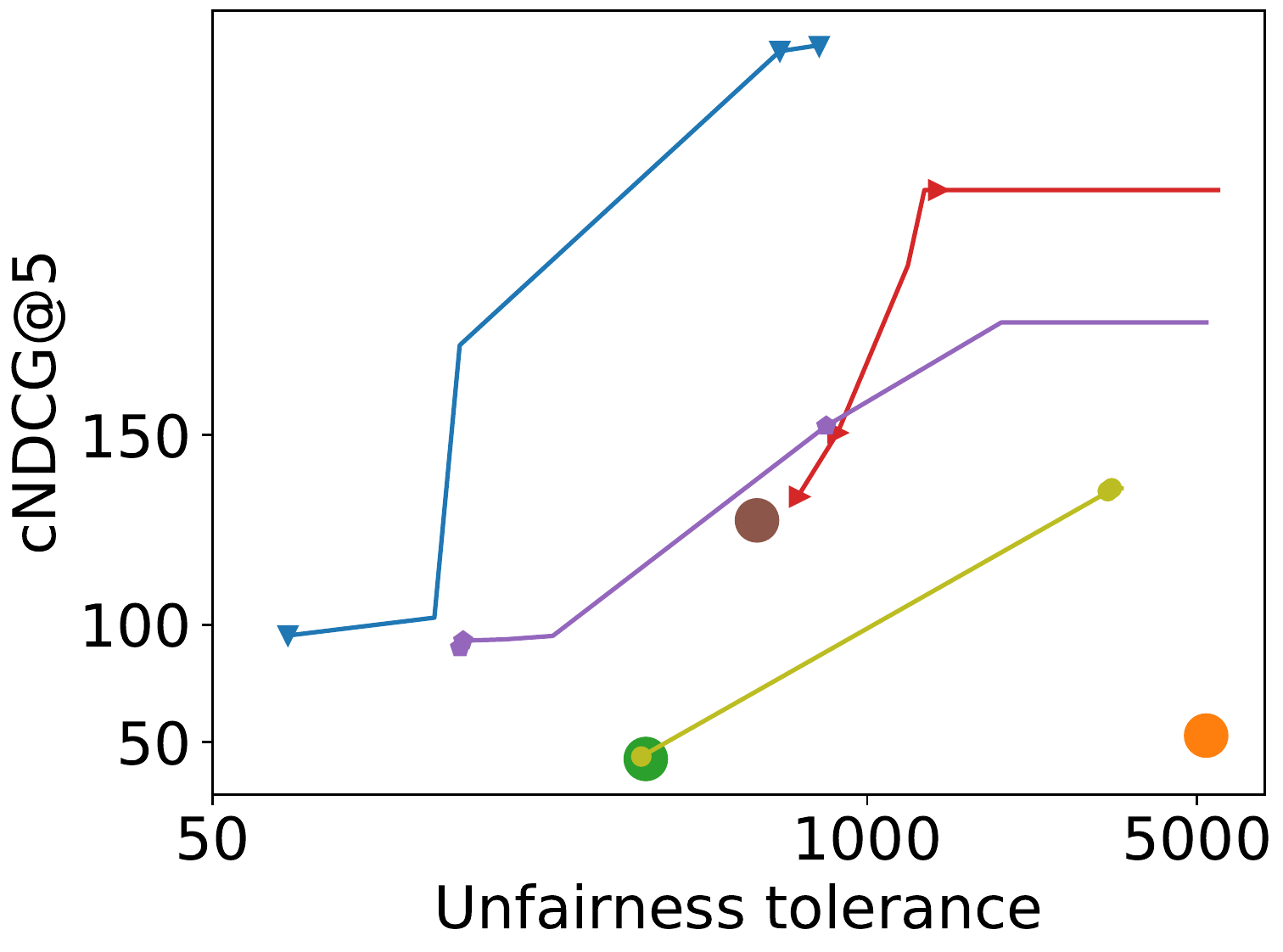}
    %\vspace{-0.5\baselineskip}
    \caption{Istella-S (online).}
    \label{fig:IsteOnline}
    \end{subfigure} \hfill
    \begin{subfigure}[]{0.10\textwidth}
    \includegraphics[scale=0.3]{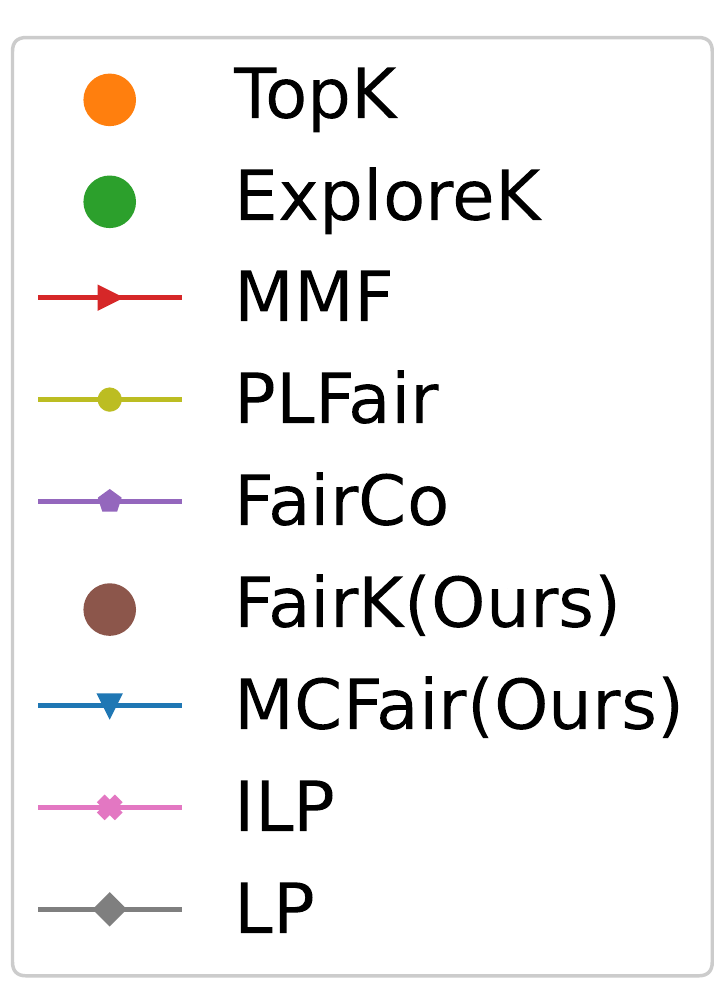}
    \end{subfigure}
    \caption{Effectiveness vs. unfairness tolerance in the post-processing setting (a,b) and the online setting (c,d). Given the same unfairness, the higher curves or points lie, the better their performances are.}
    \label{fig:balance-Post}
\vspace{-10pt}
\end{figure*}

\subsubsection{Can MCFair reach fairness in the post-processing setting?}  
\label{sec:FairCapa}
In Table \ref{tab:performance}, we compare different methods' capacity to reach fairness, where we prioritize fairness by setting $\alpha$, if available, as the its maximum  value. 
As shown in Table \ref{tab:performance}, fair ranking algorithms FairCo, LP, MCFair and FairK can significantly outperform unfair ranking algorithms in term of unfairness mitigation. 
The success of MCFair and FairK validates our assumption that fairness's gradient can be directly used as ranking scores to optimize fairness. Besides,
ILP and MMF show a slightly inferior fairness capacity and PLFair can not mitigate unfairness. More detail discussion and possible reason for their poor performance is in the next section.

Aside from unfairness, we also show the cNDCG performance of different prefixes in Figure \ref{fig:NDCGcutoff} where $\alpha$ is set to the maximum for each method. In Table \ref{tab:performance},
unfair ranking method TopK can get the highest cNDCG performance as TopK only cares about effectiveness and sacrifices fairness. 
ExploreK only explores items and thus does not optimize fairness or effectiveness. 
All the fair algorithms except MMF  and PLFair have very similar cNDCG@5 on both datasets which empirically shows that cNDCG@$k_s$ (here $k_s=5$) is fixed as we derived in Eq. \ref{eq:fixedEff}. 
Although we use $\gamma=1$ for derivation in Eq. \ref{eq:fixedEff} while using $\gamma=0.995$ to calculate cNDCG@$k_s$, similar cNDCG@$k_s$ still holds. 
Despite similar cNDCG@5 and fairness, MCFair and FairK still significantly outperform other fair algorithms at the top ranks' effectiveness. 
And we believe higher performance at top ranks' is more important as users usually pay more attention (i.e., higher examining probability) to them.  

Besides fairness capacity and effectiveness, we also empirically compare the time efficiency.
In Table~\ref{tab:performance}, ILP and LP are really time-consuming, while PLFair and MMF also need more time than other algorithms. 
While the other algorithms have similar time costs. 
% \subsubsection{how is the time efficiency?}
% All methods except LP and ILP have very similar efficiency while programming methods is very slow.

\subsubsection{Can MCFair reach a better balance between fairness and effectiveness?}  
\label{sec:balancePost}
In the previous sections, we compare algorithms when they only care about fairness. 
However, such a comparison is not sufficient since different ranking systems may have different fairness and effectiveness requirements. 
To give a comprehensive comparison of different ranking methods, we compare ranking methods' effectiveness-fairness balance given different fairness requirements.

In Figure~\ref{fig:MQPost} and Figure~\ref{fig:IstePost}, we show the balance between fairness and effectiveness in the post-processing setting. 
To generate the balance curves, we incrementally sample $\alpha$ from the minimum value to the maximum value within $\alpha$'s ranges indicated in Section~\ref{sec:baselines}. 
After sampling, we carry out five independent experiments for each $\alpha$ to get the effectiveness and fairness pair based on the average performance of the five independent experiments. Then we connect different $\alpha$'s effectiveness-fairness pair to form a curve in Figure \ref{fig:balance-Post}. 
The curves start from the right to the left as  $\alpha$ increases, and we care more about fairness.
Since TopK, ExploreK and FairK don't have trade-off parameters, each one of them only have one single pair of effectiveness and fairness and their performances are shown as single points in Figure \ref{fig:balance-Post}. 
Besides, the left bottom part of curves means caring fairness only, which corresponds to the results in Table \ref{tab:performance} and Figure \ref{fig:NDCGcutoff}.  
All curves show a tradeoff between effectiveness and fairness, which means that mitigating more unfairness usually sacrifices effectiveness. 
The reason behind this tradeoff is that achieving fairness will bring constraints on optimizing effectiveness. Among all fair methods, our method MCFair significantly outperforms all other fair methods where MCFair reaches the best cNDCG given the same unfairness, i.e., MCFair's curve lies higher. ILP, MMF, and PLFair can also show the tradeoff between effectiveness and fairness, although they have poor fairness capacities (discussed in Sec~\ref{sec:FairCapa}). As for the possible reason, the integer linear programming method used by ILP may not be effective in optimizing fairness. MMF actually follows a slightly different definition of fairness which require fairness at any cutoff should be fair. As for PLFair, PLFair tries to learn the ranking score that optimizes fairness based on the feature representation (the original setting in \cite{oosterhuis2021computationally}). However, the feature representation is originally designed for relevance which makes PLFair suboptimal.

\subsection{Results in the Online Setting.}
In the above sections, we analyze the results in the post-processing setting where relevance is assumed to be given or already well-estimated. 
In this section, we will analyze the results in a more practical setting, i.e., the online setting, where ranking optimization and relevance estimation are carried out at the same time.
% \begin{figure*}[t]
%     \centering
%         \begin{subfigure}[]{0.23\textwidth}
%     \includegraphics[scale=0.3]{figures/RealworldpositionBiasSeverity_1MQ2008test_NDCG_1_cumutradeoffplot.pdf}
%         \caption{MQ2008.}
%         \label{fig:dist_a}
%     \end{subfigure}
%     \hfill
%         \begin{subfigure}[]{0.23\textwidth}
%     \includegraphics[scale=0.3]{figures/RealworldpositionBiasSeverity_1MQ2008test_NDCG_5_cumutradeoffplot.pdf}
%         \caption{MQ2008.}
%         \label{fig:dist_a}
%     \end{subfigure}\hfill
%         \begin{subfigure}[]{0.23\textwidth}
%     \includegraphics[scale=0.3]{figures/RealworldpositionBiasSeverity_1isttest_NDCG_1_cumutradeoffplot.pdf}
%         \caption{Istella-S.}
%         \label{fig:dist_a}
%     \end{subfigure}\hfill
%         \begin{subfigure}[]{0.23\textwidth}
%     \includegraphics[scale=0.3]{figures/RealworldpositionBiasSeverity_1isttest_NDCG_5_cumutradeoffplot.pdf}
%         \caption{Istella-S.}
%         \label{fig:dist_a}
%     \end{subfigure}
%     \begin{subfigure}[]{0.43\textwidth}
%     \includegraphics[scale=0.4]{figures/positionBiasSeverity_1MQ2008legend.pdf}
%         \label{fig:dist_a}
%     \end{subfigure}    
%     \caption{Effectiveness vs. unfairness in online setting. Given the same unfairness, the higher curves or points lie, the better their performance.}
%     \label{fig:balance-IN}
% \end{figure*}

\begin{figure}[t]
    % \vspace{-2pt}
    \centering
    \begin{subfigure}[b]{\columnwidth}
    \includegraphics[scale=0.43]{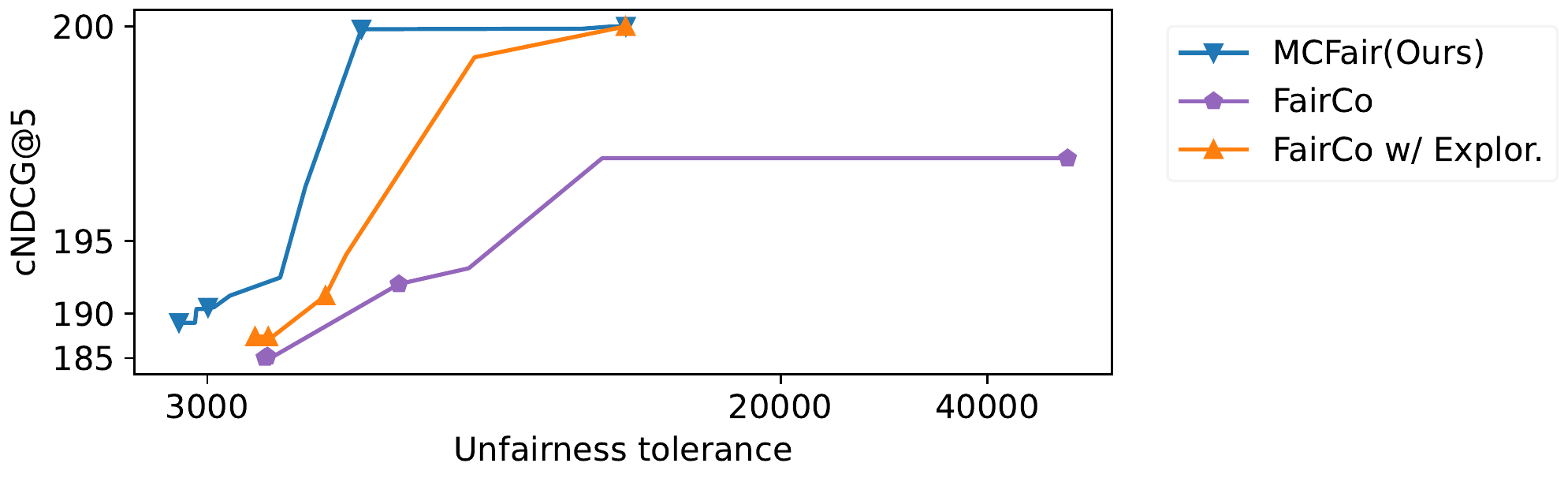}
    % \vspace{-5pt}
    % \caption{MQ2008 dataset.}
    % \label{fig:dist_a}
    \end{subfigure}
    % \begin{subfigure}[b]{0.45\textwidth}
    % \includegraphics[scale=0.40]{figures/test_NDCG_5_cumupositionBiasSeverity_1istFairCoExplorationRe.pdf}
    % \vspace{-10pt}
    % \caption{Istella-S dataset.}
    % \label{fig:dist_a}
    % \end{subfigure}
    % \vspace{-5pt}
    \caption{FairCo boosted by exploration with marginal certainty based exploration. (MQ2008)}
    \label{fig:FairCoExplo}
    \vspace{-5pt}
\end{figure}
\begin{figure}[t]
    \centering
    \begin{subfigure}[b]{\columnwidth}
    \includegraphics[scale=0.43]{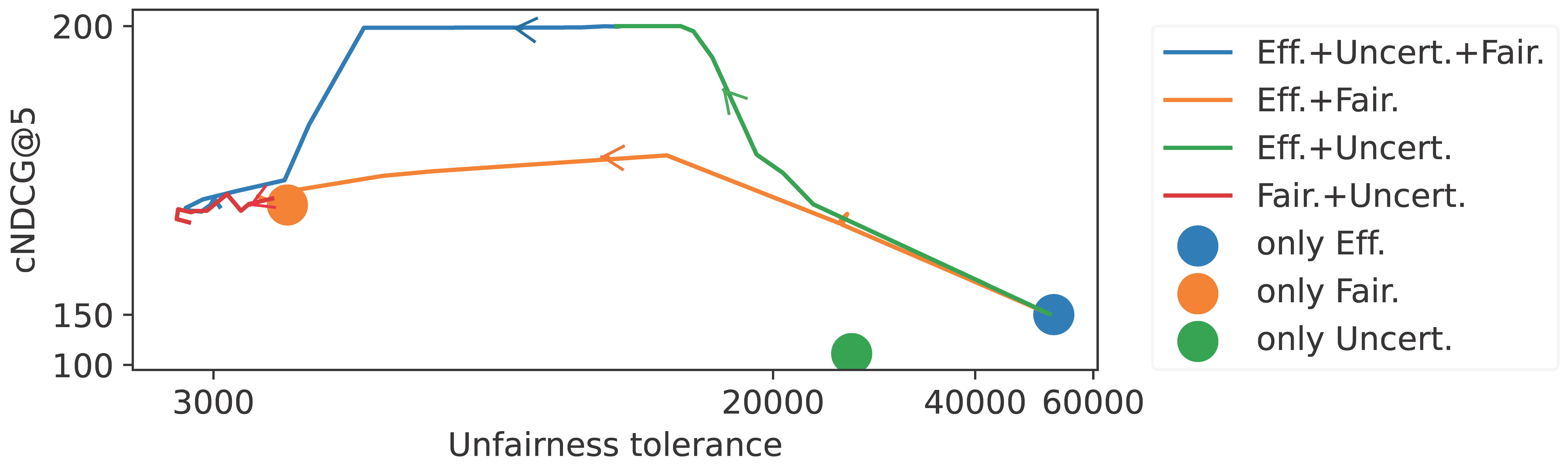}
    % \caption{MQ2008 dataset.}
    \label{fig:dist_a}
    \end{subfigure}
    % \begin{subfigure}[b]{0.45\textwidth}
    % \includegraphics[scale=0.4]{figures/test_NDCG_5_cumupositionBiasSeverity_1istAblation_tradeoffplotarrow.jpg}
    % \caption{Istella-S dataset.}
    % \label{fig:dist_a}
    % \end{subfigure}
    \vspace{-5pt}
    \caption{Ablation study of MCFair with different combinations of the three parts in Eq. \ref{eq:maginalObjWithUncer} on MQ2008. Only considering one part is shown as scatter point. When considering more than one part, we will get a curve which is generated by increasing weight of the last part. Arrow shows how curves develop when increasing the weight of the last part. For example, for Eff.+Uncert.+Fair. and Eff.+Fair., we increase the weights of Fair. i.e., fairness part, when optimizing a ranked list. With each different weight, we can get a (cNDCG@5, unfairness) pair when finishing experiments. And we connect pairs from different weights to form a curve. For Eff.+Uncert. and Fair.+Uncert., we increase the weights of uncertainty.}
    \label{fig:abalation}
    % \vspace{-2pt}
\end{figure}

\subsubsection{Can MCFair work in the online setting?}  
To study this problem, we show the balance between effectiveness and fairness of the online setting  in Fig. \ref{fig:MQOnline} and Fig. ~\ref{fig:IsteOnline}.
Similar to the post-processing setting, our method MCFair still outperforms all other baselines in the online setting. Since most of the results are similar to the post-processing setting in \S\ref{sec:balancePost}, we only discuss the difference.
% Please refer to Section~\ref{sec:balancePost} for how we generate the balance graph. 
% Experimental results in the online setting and the post-processing setting are pretty different. 
In Fig. \ref{fig:MQOnline} and Fig. ~\ref{fig:IsteOnline}, TopK  cannot reach the highest effectiveness and FairK also can not reach the lowest unfairness in the  online setting, which is different from the post-processing setting.  
We think the reason for the difference is that they naively trust the uncertain relevance estimation, which makes effectiveness optimization and fairness optimization fail (more discussion in \S\ref{sec:ablation}).

\subsubsection{Can marginal certainty help boost existing fair methods' performance?}  
\label{sec:FairCohelp}
In this section, we investigate whether the marginal certainty-based exploration can boost existing fair methods. 
We mainly focus on how to boost FairCo and leave how to boost other fair methods for future study. 
Specifically, we directly add $\reallywidehat{MC}(d)$ (see Eq.~\ref{eq:MC}), the marginal certainty, to FairCo's ranking score, referred to as \textit{FairCo w/ Explor.}  
As shown in Figure~\ref{fig:FairCoExplo}, FairCo w/ Explor. outperforms FairCo as it reaches better cNDCG given the same unfairness. 
FairCo w/ Explor.'s better performance shows marginal certainty can effectively boost FairCo. 
Our method MCFair still significantly outperforms FairCo w/ Explor.

% \begin{figure}[t]
%     \centering
%         \begin{subfigure}[]{0.5\textwidth}
% \includegraphics[scale=0.45]{figures/positionBiasSeverity_1MQ2008NDCGcutoffcumu.pdf}
%         \caption{MQ2008.}
%         \label{fig:dist_a}
%     \end{subfigure}
%         \begin{subfigure}[]{0.5\textwidth}
%     \includegraphics[scale=0.45]{figures/positionBiasSeverity_1istNDCGcutoffcumu.pdf}
%         \caption{Istella-S.}
%         \label{fig:dist_a}
%     \end{subfigure}
%     \caption{Post-processing results.}
%     \label{fig:NDCGcutoff}
% \end{figure}

% \subsubsection{How does the exploration influence?}
% \begin{figure}
%     \centering
%         \begin{subfigure}[]{0.5\textwidth}
% \includegraphics[scale=0.45]{figures/RealworldpositionBiasSeverity_1MQ2008test_NDCG_5_cumuExploraDegree.pdf}
%         \caption{MQ2008.}
%         \label{fig:dist_a}
%     \end{subfigure}
%         \begin{subfigure}[]{0.5\textwidth}
%     \includegraphics[scale=0.45]{figures/RealworldpositionBiasSeverity_1istella-stest_NDCG_5_cumuExploraDegree.pdf}
%         \caption{Istella-S.}
%         \label{fig:dist_a}
%     \end{subfigure}
%     \caption{FairCo boosted by exploration with marginal uncertainty.}
%     \label{fig:explorationDegree}
% \end{figure}

% As shown in Figure~\ref{fig:explorationDegree}, more exploration can help to gain better performance.
\subsubsection{Ablation study. } 
\label{sec:ablation}
In this section, we conduct an ablation study to evaluate the significance of each part of  MCFair. Since the ranking score of MCFair $\reallywidehat{ug}(d)$ in Eq. \ref{eq:MarginalObjectiveUnc} has three parts, corresponding to effectiveness, fairness, and uncertainty respectively, there are a total of seven (${3\choose 1}+{3\choose 2}+{3\choose 3}=7$) combinations that need to be evaluated. 
The ablation results of each combination are shown in Figure~\ref{fig:abalation}. 
For the ablation results, considering more than two parts are shown as balance curves while considering only one part is shown as single points. In Figure~\ref{fig:abalation}, there is a very interesting cycle formed by the curves. 
In the cycle, considering all three parts, i.e., Eff.+Fair.+Uncer., outperform all other combinations since Eff.+Fair.+Uncer. can reach better effectiveness given the same unfairness.

\section{Conclusions}
In this work, we study the critical problem of relevance-fairness balance in online ranking settings.
We propose a novel Marginal-Certainty-aware Fair Ranking algorithm named MCFair. 
MCFair jointly optimizes fairness and user utility while relevance estimation is constantly updated in an online manner.
With extensive experiments on semi-synthesized datasets, MCFair shows its superior performance compared to other fair ranking algorithms.
\section*{Acknowledgements}
This work was supported  by the School of Computing, University of Utah. Any opinions, findings and conclusions or recommendations expressed in this material are those of the authors and do not necessarily reflect those of the sponsor.

\pagebreak
% \clearpage
\bibliographystyle{ACM-Reference-Format}
\bibliography{mybib}

% \pagebreak

\end{document}